\documentclass[aps, pre, twocolumn, showpacs]{revtex4}
\usepackage{amsmath, amsfonts, amssymb}
\usepackage{bm}
\usepackage{hyperref}
\usepackage{graphicx}

\def\Re {\mbox{Re}}
\def\Im {\mbox{Im}}

\newcommand{\di}{\mathrm{d}}
\newcommand{\ii}{i}

\renewcommand{\vec}[1]{\mathbf{#1}}
\newcommand{\ket}[1]{|#1\rangle}

\renewcommand{\vr}{{\vec{r}}}

\begin{document}

\title{Tunneling of anyonic Majorana excitations in topological superconductors}
\author{Meng Cheng$^1$}
\author{Roman M. Lutchyn$^{1,2}$}
\author{Victor Galitski$^{1,2}$}
\author{S. Das Sarma$^{1,2}$}

\affiliation{$^{1}$Condensed Matter Theory Center, Department of Physics,
             University of Maryland, College Park, Maryland 20742, USA\\
$^{2}$Joint Quantum Institute, Department of Physics,
             University of Maryland, College Park, Maryland 20742, USA}

\date{\today}

\begin{abstract}
We consider topological superconductors and topological insulator/superconductor structures in the presence of 
multiple static vortices that host Majorana modes and focus on the Majorana tunneling processes between vortices.
It is shown that these tunnelings generally lift the degeneracy of the many-body ground state in a non-universal way, 
sensitive to microscopic details at the smallest length-scales determined by the underlying physical problem. 
We also discuss an explicit realization of the Jackiw-Rossi zero-mode in a topological insulator/superconductor 
structure with zero chemical potential. In this case, the exact degeneracy of the many-anyon ground state is protected 
by an additional chiral symmetry and can be linked to the rigorous index theorem. However, the existence 
of a non-zero chemical potential, as expected in realistic solid state structures, breaks chiral symmetry and removes 
protection, which leads to the degeneracy being lifted. Finally, we discuss the implications of our results for the collective states of many-anyon systems. 
We argue  that quantum dynamics of vortices in 
realistic systems is generally important and may give rise to effective time-dependent gauge factors that enter interaction 
terms between Majorana modes in many-anyon systems.
\end{abstract}


\maketitle

\section{Introduction}

Topological quantum computation hinges on the existence of non-Abelian excitations,
which arise  in certain topological phases of matter~\cite{nayak_RevModPhys'08}. The first
example of such a state is Fractional Quantum Hall (FQH) state  with $\nu=5/2$
filling fraction.  This state is believed to be described by the Moore-Read Pfaffian
wave function~\cite{Moore_NPB91} which supports non-Abelian
excitations~\cite{Nayak_NPB96}. It was shown later that the Moore-Read Pfaffian
wave function of composite fermions is related to the BCS wave function with
p-wave pairing~\cite{Moore_NPB91, Greiter_NPB92, read_prb'00,Volovik_book} establishing the
connection between FQH state $\nu=5/2$ and $p_x+i p_y$ superconductors. Certain
vortex excitations in chiral p-wave superconductors carry zero energy modes and
obey non-Abelian statistics~\cite{read_prb'00, Volovik_JETP'99, Ivanov_PRL'01}.
These topologically protected zero modes can be occupied by Majorana fermions
and are responsible for topological ground state degeneracy. Namely, $2n$
vortices with zero modes residing in vortex cores span $2^{n-1}$-dimensional
Hilbert space. Non-Abelian statistics of vortices can also be derived within
this framework~\cite{Ivanov_PRL'01}. Making use of their intrinsic non-local
quantum entanglement, vortices carrying Majorana modes can be exploited to
realize topological qubits which are inherently decoherence-free and are
protected against smooth local perturbations, thus providing a very appealing
platform for fault-tolerant topological quantum
computation~\cite{Kitaev_AP03,dassarma_prl'05,tewari_prl'2007}. 

From the perspective of experimental realization of topological phases, 
there is some preliminary evidence that $\nu=5/2$ FQH state may have non-Abelian 
excitations~\cite{Radu_Sci08, Willet_PNAS09, Nayak_PRB'09}. Spin-triplet $p_x+ip_y$ 
pairing superfluidity/superconductivity are believed to occur in $A$-phase of superfluid 
$^3\text{He}$~\cite{Kopnin_PRB'91, Tsutsumi_PRL08} and strontium 
ruthenates~\cite{Mackenzie_RevModPhys'03, Kapitulnik_PRL'06, lutchyn_prb'08, Lutchyn_prb'09, Kallin} 
in which half-quantum vortices are non-Abelian~\cite{Ivanov_PRL'01, SDS_PRB06}. There are also proposals 
to realize chiral p-wave superfluids in ultracold atom 
systems~\cite{Gurarie20072, Mizushima_PRL08, zhang_prl'08, Sato_PRB09,Nishida_AnnPhys09,  Cooper_prl'09}. 
The theoretical description of these systems essentially falls into the category of a spinless $p_x+ip_y$ superconductor. 
Apart from these examples, there also exist a number of proposals involving various heterostructures of a three dimensional topological 
insulator (TI) and a superconductor (SC)~\cite{Fu_PRL08, Linder_PRL10}, semiconductor and superconductor~\cite{Sau_PRL10, Alicea_PRB10} and superconductor and ferromagnet~\cite{Lee_preprint09, Qi'10}. These systems seem to be more experimental accessible. We also note that there are proposals to realize Majorana fermions in one-dimensional systems~\cite{Kitaev_Majorana, Fu_PRB2009, Wimmer_arxiv10, Lutchyn_preprint10, Oreg_2010} and on 
the surface of a three dimensional $\mathbb{Z}_2$ topological superfluid/superconductor~\cite{Roy_preprint08, Chung_PRL09, Qi_PRL09, Volovik_JETP09}.

We note here that the emergent Majorana excitations in these physical systems are closely related to zero-modes that
have been long known in the context of high-energy physics, where chiral fermions in the presence of topological defects (domain walls,
vortices, etc.) give rise to massless excitations within the topological defects~\cite{tHooft}. One such example is a Jackiw-Rossi zero-mode
that was predicted to appear in a vortex-chiral-fermion system.  An important ingredient of the original Jackiw-Rossi model~\cite{Jackiw_NPB81} 
is the existence of a conserved chiral current, which allows one to enumerate zero-energy modes according to their chirality. It was subsequently established by Weinberg~\cite{Weinberg_PRD81} that for the Dirac operator the difference between the number of zero modes of opposite chirality is given by the winding number of the superconducting order parameter phase. Thus, the conservation of the chiral current enables the use of a powerful Atiyah-Singer-type index theorem~\cite{Atiyah_AnnMath68, Jackiw_private}, which relates
a number of zero modes to the total topological charge of the vortex configuration. In particular, the theorem ensures that
the exact degeneracy of the many-body ground state is preserved and no tunneling process can possibly lift it.
As discussed in Refs.~[\onlinecite{Fu_PRL08, Chamon_PRB10}], a Hamiltonian considered by Jackiw-Rossi can be realized in a TI/superconductor heterojunction. However, as shown below a realistic solid-state structure of this type is generally described by a low-energy theory, which has the exact 
chiral symmetry only if the chemical potential of excitations is exactly zero. Due to the non-trivial way the chemical potential enters the BdG equations, there is no conserved chiral current, and one can not enumerate zero modes by the their chirality anymore. Therefore, the connection between the analytical and topological index established in Ref.~\cite{Weinberg_PRD81} does not apply in this case, and intervortex tunneling leads to lifting of the ground-state degeneracy of a many-anyon system for any finite chemical potential. Clarification of the applicability of the index theorem to the non-relativistic topological superconductors is the main result of the paper. 
 
Understanding the fate of ground-state degeneracy of many-anyon system in realistic solid-state structures is a difficult problem of fundamental
importance and of relevance to practical realization of topological quantum computing. In this paper we address one mechanism that may lift the ground state degeneracy associated with the tunneling processes between spatially separated vortices. The presence of the bulk gap protects ground state degeneracy from thermal fluctuations at low temperature leaving out only processes  of Majorana fermion quantum tunneling between vortices. Generic features of tunneling of topological charges have been explored recently~\cite{Bonderson_PRL09}. The lifting of ground state degeneracy due to intervortex tunneling for a pair of vortices have been studied numerically for $\nu=5/2$ quantum Hall state~\cite{Tserkovnyak_PRL03,Baraban_PRL09}, $p_x+ip_y$ superconductor~\cite{Kraus_PRB09} and Kitaev's honeycomb lattice model~\cite{Lahtinen_AnnPhys08}. Analytical calculation has been carried out for the model of spinless $p_x+ip_y$ superconductors~\cite{Cheng_PRL09}. 

Generally energy splitting due to intervortex tunneling is determined by the wave function overlap of localized Majorana bound states. In this paper we calculate the splitting for both spinless $p_x+ip_y$ superconductor and a model of Dirac fermions interacting with the scalar superconducting pairing potential realized in a TI/SC heterostructure. In both cases, besides the expected exponential decay behavior, it is found that the prefactor exhibits an oscillatory behavior with the intervortex distance which originates from the interference of different bound state wave functions oscillating with the Fermi wave length. This is generic situation for weak coupling superconductors where the Fermi energy $E_F$ is much larger than the superconducting gap $\Delta$. In this paper, we also consider several cases where the Fermi wavelength is much larger than the coherence length. This scenario is relevant, for example, for TI/SC heterostructure as well as some other systems involving the proximity-induced superconductivity.  When chemical potential is tuned to the Dirac point ($\mu \rightarrow 0$), we find indeed that the splitting in TI/SC heterostructure vanishes. This fact can be attributed to an additional symmetry possessed by the system at $\mu=0$ - the chiral symmetry as discussed above.  

The paper is organized as follows. In Sec. \ref{sec:majorana_state} we review the Bogoliubov-de Gennes equations for spinless $p_x+ip_y$ superconductor as well as TI/SC heterostructure and show that there are Majorana zero energy  solutions localized at the vortex core. Then using these bound state wave functions, we calculate energy splitting of zero-energy states due to intervortex tunneling in Sec. \ref{sec:splitting}.  We present our main results in Sec.~\ref{sec:index} by interpreting the explicit splitting calculations presented in the previous section from the perspective of the index theorem which establishes the relation between zero-energy modes and topological index of the order parameter. Implications of our result for topological quantum computation and interacting many anyons system are discussed in Sec.~\ref{sec:collective}. 

\section{Zero energy bound states in superconducting vortex cores}
\label{sec:majorana_state}
In this Section, we review the analytic solution of the Bogoliubov-de Gennes equation for a zero-energy Majorana mode 
in a $p+ip$ superconductor and at a topological insulator/superconductor interface. The main results here are Eqs.~(\ref{onevortex})
and (\ref{eq:zeromodetisc2}) and they are used further in Sec.~III to calculate the energy-level splitting due to tunneling. 

\subsection{Bound states in $p_x+ip_y$ superconductors}
We start with the mean-field Hamiltonian for spinless $p_x+i p_y$ superconductor
\begin{multline}
  \mathcal{H}_{\text{BCS}}=\int\di^2\vr\,\hat{\psi}^\dag(\vr)\left(-\frac{\nabla^2}{2m}-\mu\right)\hat{\psi}(\vr)+\\
  \frac{1}{2}\int\di^2\vr\di^2\vr'\,\left[\hat{\psi}^\dag(\vr)\Delta(\vr,\vr')\hat{\psi}^\dag(\vr')+\text{h.c.}\right],
  \label{eq:BCSH}
\end{multline}
where the gap operator $\hat{\Delta}(\vr,\vr')$ is  given by~\cite{Stern_prb'04}
\begin{equation}
 \Delta(\vr,\vr')=\frac{1}{k_F}\Delta\left(\frac{\vr+\vr'}{2}\right)(\partial_{x'}+i\partial_{y'})\delta(\vr-\vr').
  \label{eq:pipgap}
\end{equation}
This Hamiltonian can be derived as a continuum limit of a lattice model of spinless fermions~\cite{Cheng_PRB2010}. To diagonalize this Hamiltonian we perform Bogoliubov transform $\hat{\psi}(\vec{r})=\sum_n \big[\hat{\gamma}_nu_n(\vec{r})+\hat{\gamma}_n^\dag v_n^\ast(\vec{r})\big]$ where $n$ labels different quasiparticle eigenstates. Canonical commutation relation $[\mathcal{H}_\text{BCS},\hat{\gamma}_n]=-E_n\hat{\gamma}_n$ yields corresponding Bogoliubov-de Gennes(BdG) equation
\begin{align}
\mathcal{H}_{\mathrm{BdG}}
\begin{pmatrix}
u(\bm r)\\
v(\bm r)
\end{pmatrix}=E\begin{pmatrix}
u(\bm r)\\
v(\bm r)
\end{pmatrix}, \label{eq:bdg}
\end{align}
where BdG Hamiltonian reads
\begin{align}
\!\!\mathcal{H}_{\mathrm{BdG}}\!\!=\!\! \begin{pmatrix}
  {\displaystyle \!-\!\frac{\nabla^2}{2m}\!-\!\mu} \!&\! {\displaystyle\frac{1}{k_F}\!\{\Delta(\vec{r}), \partial_x\!+\!i\partial_y\}}\\
{\displaystyle \!-\!\frac{1}{k_F}\!\{\Delta^\ast(\vec{r}), \partial_x \! - \! i\partial_y\}} \!&\! {\displaystyle\frac{\nabla^2}{2m}\!+\!\mu}
\!\end{pmatrix}\!\!
\label{eq:Hbdg}
\end{align}
with anti-commutator being defined as $\{a,b\}=(ab+ba)/2$.

Before discussing the zero-energy solutions of BdG equations, it is instructive to review the symmetries of the Bogoliubov-de-Gennes Hamiltonian~\eqref{eq:Hbdg}. Particle-hole symmetry of BdG Hamiltonian follows from $\{\Xi,\mathcal{H}_\text{BdG}\}=0$ where $\Xi=\tau_xK$ with $K$ being complex conjugation operator and $\tau_x$ being Pauli matrix in Nambu(particle-hole) space~\cite{Gurarie_prb'07}. Besides particle-hole symmetry BdG Hamiltonian \eqref{eq:Hbdg} has no other generic symmetries. Thus, it is a typical example of the symmetry class D in the general classification scheme of topological insulators and superconductors~\cite{Schnyder_PRB08, Kitaev2009}. The direct consequence of particle-hole symmetry is that the eigenstates of $H_{\rm BdG}$\eqref{eq:Hbdg} come in pairs, with opposite eigenenergies. That is, if $\Psi=(u_n,v_n)^T$ is a solution of Eq.~(\ref{eq:bdg}) with eigenvalue $E_n$, then $\Xi\Psi=(v_n^\ast,u_n^\ast)^T$ must be a solution with the eigenvalue $(-E_n)$. Particularly, a non-degenerate zero energy state must obey the following constraint: $\Xi\Psi=\lambda\Psi$. Because $\Xi^2=1$ which implies $|\lambda|=1$, $\lambda$ must be a pure phase $\lambda=e^{i\theta}$. We can make a global gauge transformation and introduce Nambu spinors as $\Psi'=\exp(-\ii \theta/2)\Psi$ then $\Xi\Psi'=\Psi'$. Thus, a non-degenerate zero energy state should always satisfy $u^\ast=v$. The corresponding quasiparticle operator 
\[
\hat{\gamma}^\dag= \int \di^2\vec{r}\,\big[u(\vr)\hat{\psi}^\dag(\vec{r})+v(r)\hat{\psi}(\vec{r})\big]
\]
is then self-Hermitian obeying $\hat{\gamma}=\hat{\gamma}^\dag$, \emph{i.e.} $\hat \gamma$ is a Majorana fermion operator. 

We will now show that such zero energy states appear in the cores of vortices in chiral p-wave superconductors.  The localized states in the vortex cores are known as Caroli-de-Gennes-Matricon states (CdGM)~\cite{Caroli_PL'64}.  In conventional s-wave superconductors all CdGM states have non-zero energies~\cite{Volovik_JETP'99}. However, due to the chirality of the order parameter $p_x+ip_y$ superconductors can host zero-energy bound states~\cite{Kopnin_PRB'91, Volovik_JETP'99,tewari_prl'07,Gurarie_prb'07, Mizushima_PRA10}.  The non-degenerate zero-energy bound states are topologically protected by the particle-hole symmetry.  The existence of the zero energy solution in the vortices of the chiral p-wave superconductors can be demonstrated explicitly by solving BdG equations~\cite{Gurarie_prb'07}. Similar to the s-wave superconductors~\cite{Caroli_PL'64,bardeen_prb'69}, a vortex with vorticity $l$(i.e. $l$ flux quanta $hc/2e$ is trapped) can be modeled as 
$\Delta(\vr)=f(r)e^{il\varphi}$
  %\label{eq:vortexgap}
  %\end{equation}
where $\varphi$ is the phase of the order parameter and $f(r)$ is the vortex profile which can be well approximated by $f(r)=\Delta_0\tanh(r/\xi)$~\cite{bardeen_prb'69}.  Here $\Delta_0$ is the mean-field value of superconducting order parameter and $\xi=v_F/\Delta_0$ is coherence length. We will mainly focus on the case $l=1$. Taking advantage of rotational symmetry, BdG equation can be decoupled into angular momentum channels. The wave function can be written as $\Psi_m(\vr)=e^{im\varphi}(e^{i\varphi}u_m(r),e^{-i\varphi}v_m(r))$. As argued above, a non-degenerate zero mode requires $\Xi\Psi\propto\Psi$ which can only be satisfied for $m=0$. The radial part of the BdG equations in $m=0$ channel then reads
  \begin{widetext}
  \begin{equation}
	\begin{pmatrix}
	  -\frac{1}{2m}(\partial_r^2+\frac{1}{r}\partial_r-\frac{1}{r^2})-\mu & \frac{1}{k_F}\left[f(r)(\partial_r+\frac{1}{2r})+\frac{f'(r)}{2}\right]\\
  -\frac{1}{k_F}f(r)\left[(\partial_r+\frac{1}{2r})+\frac{f'(r)}{2}\right] & \frac{1}{2m}(\partial_r^2+\frac{1}{r}\partial_r-\frac{1}{r^2})+\mu
	\end{pmatrix}\begin{pmatrix}
	u_0(r)\\
	v_0(r)
  \end{pmatrix}=0.
	\label{eq:rbdg}
  \end{equation}
\end{widetext}
Given that the radial part of the BdG equation  \eqref{eq:rbdg} is real, one can choose $u_0(r)$ and $v_0(r)$ to be real. Then the condition $\Xi\Psi_0=\Psi_0$ reduces to $v_0=\lambda u_0$ with $\lambda=\pm 1$. Using this constraint, the differential equation for $u_0$ becomes: 
\[
\left\{\left(\partial_r^2\!+\!\frac{1}{r}\partial_r\!-\!\frac{1}{r^2}\right)\!-\!2m\mu \!-\! \frac{2\lambda}{v_F}\left[f\left(\partial_r\!+\!\frac{1}{2r}\right)\!+\!\frac{f'}{2}\right]\right\}u_0\!=\!0. 
\]
One can seek the solution of the above equation in the form $u(r)=\chi(r)\exp\left[\lambda\int_0^r\di r'\,f(r')\right]$ leading to  
\begin{equation}
  \chi''+\frac{\chi'}{r}+\left(2m\mu-\frac{f^2}{v_F^2}-\frac{1}{r^2}\right)\chi=0.
  \label{eq:radialfinal}
\end{equation}
Here the profile $f(r)=\Delta_0\tanh(r/\xi)$ vanishes at the origin and reaches $\Delta_0$ away from vortex core region. For our purpose, it's sufficient to consider the behavior of solution outside the core region where $f(r)$ is equal to its asymptotic bulk value $\Delta_0$. It is obvious now that $\lambda=-1$ yields the only normalizable solution. 

When $\Delta_0^2< 2m\mu v_F^2$ which is the case for weak-coupling BCS superconductors, Eq.\eqref{eq:radialfinal} becomes first order Bessel equation. Thus, the solution is given by Bessel function of the first kind $J_n(x)$:
\begin{equation}
\chi(r)=\mathcal{N}_1J_1(r\sqrt{2m\mu-\Delta_0^2/v_F^2}),
\label{eq:chi1}
\end{equation}
where $\mathcal{N}_1$ is the normalization constant determined by the following equation $4\pi\int r\di r\,|u_0(r)|^2=1$.   
%Besides, spectrum of low-lying excited states can be found with quasi-classical approximation as $E_m=-\omega_0 m$~\cite{Mizushima_arxiv2010}. Here $\omega_0\approx \Delta_0^2/\varepsilon_F$ characterizes seperation between these energy levels. Majorana bound state is protected by this mini-gap $\omega_0$.
In the  opposite limit $\Delta_0^2> 2m\mu v_F^2$, the solution of Eq.~\eqref{eq:radialfinal} is given by first order imaginary Bessel function: 
\begin{equation}
\chi(r)=\mathcal{N}_2I_1(r\sqrt{\Delta_0^2/v_F^2-2m\mu}). 
\label{eq:chi2}
\end{equation}
The function $I_n(r)$ diverges when $r\rightarrow \infty$. But the radial wave function $u_0(r)$ remains bounded as long as $\mu>0$. This is consistent with the fact that $\mu=0$ separates Abelian topological phase ($\mu<0$) and non-Abelian phase($\mu>0$)~\cite{read_prb'00}. The critical value $\Delta_0^2=2m\mu v_F^2$ corresponds to closing of the bulk gap. At this point the notion of a localized bound state becomes meaningless. Indeed, at this point the solution of  Eq.~\eqref{eq:radialfinal} scales as $\chi(r)\propto r$. 

We summarize this section by providing an explicit expression for zero energy eigenfunction:
\begin{equation}
  \!\Psi_0(\vec{r})=\chi(r)\!\exp\!\left[\! i\left(\varphi-\frac \pi 2\right) \tau_z\!-\!\frac{1}{v_F}\!\!\int_0^r\! \di r'f(r')\!\right]\!, 
\label{onevortex}
\end{equation}
where $\chi(r)$ is given by Eq.\eqref{eq:chi1} for $\Delta_0^2< 2m\mu v_F^2$ and Eq.\eqref{eq:chi2} for $\Delta_0^2> 2m\mu v_F^2$.

Using the zero energy solution obtained for one vortex one can be easily write down wave function for multiple vortices spatially separated so that tunneling effects can be ignored. Assume there are $2N$ vortices pinned at positions $\vec{R}_i\,,i=1,\dots,2N$. The superconducting order parameter can be represented as 
\begin{equation}
	\Delta(\vec{r})=\prod_{i=1}^{2N}f(\vec{r-R_i})\exp\big[\ii\sum_i\varphi_i(\vec{r})\big],
	\label{eq:mulvortices}
\end{equation}
where $\varphi_i(\vec{r})=\mathrm{arg}(\vec{r}-\vec{R}_i)$. Near the $k$-th vortex core, the phase of the order parameter is well approximated by $\varphi_k(\vec{r})+\Omega_k$ with $\Omega_k=\sum_{i\neq k}\varphi_i(\vec{R}_k)$ which is accurate in the limit of large intervortex seperation. Then in the vicinity of $k$-th vortex core, a zero energy bound state can be found~\cite{Stern_prb'04}:
\begin{align}
  \Psi_k(\vec{r})&=e^{-i\tau_z\frac \pi 2}\chi(r_k)\exp\left[-\frac{1}{v_F}\int_0^{r_k} \di r'f(r')\right]\nonumber\\
& \times \exp\left[ i \left(\varphi_k+\frac{\Omega_k}{2}\right)\tau_z\right].
\label{eq:zeromode2}
\end{align}
where $r_k=|\vec{r}-\vec{R}_k|$. Correspondingly, there are $2N$ Majorana fermion modes localized in the vortex cores. They can be combined pairwise to form $N$ Dirac fermions. Specifically, two Majorana fermions $\hat{\gamma}_i$ and $\hat{\gamma}_j$ localized in $i$-th and $j$-th vortex cores, respectively, are combined into a Dirac fermion:
\begin{equation}
  \hat{c}^\dag=\frac{1}{\sqrt{2}}(\hat{\gamma}_i+i\hat{\gamma}_j).
  \label{eq:diracfrommajorana}
\end{equation}
These $N$ Dirac fermions can be occupied or unoccupied, allowing for enumeration of all degenerate ground states~\cite{nayak_RevModPhys'08}.
\subsection{Bound states in the Dirac fermion model coupled with s-wave superconducting scalar field.}

We now discuss the zero energy bound states emerging in the model of Dirac fermions interacting with the superconducting pairing potential. This model is realized at the interface of a 3D strong topological insulator having an odd number of Dirac cones per surface and an s-wave superconductor~\cite{Fu_PRL08}.  
%Recently there have been several appealing proposals~\cite{Fu_PRL08} of realizing Majorana zero modes in heterostructure basically composed of an insulator with certain non-trivial band structure(usually in the presence of spin-orbit coupling) in close contact with an ordinary s-wave superconductor. 
Due to the proximity effect an interesting topological state is formed at the 2D interface between the insulator and superconductor. We will now discuss the emergence of Majorana zero energy states at the TI/SC heterostructure~\cite{Fu_PRL08}. This model was also considered earlier in the high-energy context by Jackiw and Rossi~\cite{Jackiw_NPB81}. 

Three dimensional time-reversal invariant topological insulators are characterized by an odd number of Dirac cones enclosed by Fermi surface~\cite{Fu_PRL07, Roy_PRB09, Moore_PRB07}. The metallic surface state is described by the Dirac Hamiltonian. The non-trivial $\mathbb{Z}_2$ topological invariant ensures the stability of metallic surface states against perturbations which preserve time-reversal symmetry. %In the simplest case where only one Dirac cone is present at the surface, surface of 3D topological insulator can be described by a Dirac Hamiltonian in the simplest case. 
When chemical potential $\mu$ is close to the Dirac point the TI/SC heterostructure can be modeled as~\cite{Fu_PRL08, Stanescu'10}:
\begin{equation}
  \mathcal{H}=\hat{\psi}^\dag(v\vec{\sigma}\cdot\vec{p}-\mu)\hat{\psi}+\Delta\hat{\psi}^\dag_{\uparrow}\hat{\psi}^\dag_{\downarrow}+\text{h.c},
	\label{eq:tiscH}
\end{equation}
where $\psi=(\psi_\uparrow,\psi_\downarrow)^T$ and $v$ is the Fermi velocity at Dirac point. The Bogoliubov-de Gennes equations are given by:
\begin{align}
	&\mathcal{H}_\text{BdG}\Psi(\vr)=E\Psi(\vr)\\
	\mathcal{H}_\text{BdG}&=\begin{pmatrix}
		\vec{\sigma}\cdot\vec{p}-\mu & \Delta\\
		\Delta^* & -\vec{\sigma}\cdot\vec{p}+\mu
	\end{pmatrix},
	\label{eq:tiscBdG}
\end{align}
where $\Psi(\vr)$ is the Nambu spinor defined as  $\Psi=(u_\uparrow, u_\downarrow, v_\downarrow, -v_\uparrow)^T$. At $\mu=0$ the BdG Hamiltonian above can be conveniently written in terms of the Dirac matrices: 
\begin{align}\label{eq:Dirac}
H_{\rm BdG}=\sum_{a=1,2}\left( \gamma_a p_a +\Gamma_a n_a \right).
\end{align} 
Here $\gamma_a$ and $\Gamma_a$ are $4 \times 4$ Dirac matrices defined as $\gamma_1=\sigma_x\tau_z, \gamma_2=\sigma_y\tau_z$, and $\Gamma_1=\tau_x,\Gamma_2=\tau_y$ and $\vec{n}=(\Re\Delta,-\Im\Delta)$. One can check that these matrices satisfy the following properties: 
$\{ \gamma_{a}, \gamma_b \}=\{\Gamma_a, \Gamma_b \}=\delta_{ab}$ and $\{\Gamma_a, \gamma_b \}=0$. The fifth Dirac matrix $\gamma_5$ is given by $\gamma_5=-\gamma_1\gamma_2 \Gamma_1 \Gamma_2=\tau_z \sigma_z$. 

As in the case of spinless $p_x+ip_y$ case, we first discuss the symmetries of Eq.\eqref{eq:tiscBdG}. The particle-hole symmetry is now $\Xi=\sigma_y\tau_yK$ where $\vec{\tau}$ are Pauli matrices operating in Nambu (particle-hole) space. The difference with the previous case is the presence of time-reversal symmetry: $\Theta=i\sigma_yK,\, [\Theta, \mathcal{H}_\text{BdG}]=0$ in this model. Moreover, when $\mu=0$ there is additional chiral symmetry in the model which can be expressed as $\{\gamma^5,\mathcal{H}_\text{BdG}\}=0$. We will see that the chiral symmetry has important implications for degeneracy splitting. Bogoliubov quasiparticles are defined from solutions of BdG equations as 
\begin{equation}
  \hat{\gamma}^\dag=\int\di^2\vr\,\sum_\sigma u_\sigma(\vr)\hat{\psi}^\dag_\sigma(\vr)+v_\sigma(\vr)\hat{\psi}_\sigma(\vr).
	\label{eq:qptisc}
\end{equation}
If we require $\hat{\gamma}$ to be a Majorana fermion, {\it i.e.} $\hat{\gamma}=\hat{\gamma}^\dag$, the necessary and sufficient condition is $v_\sigma=u^*_\sigma$ up to a global phase. 

A vortex with vorticity $l$ can be introduced in the order parameter as $\Delta(\vr)=f(r)e^{il\varphi}$. Rotational symmetry allows decomposition of solutions into different angular momentum channels: 
\[
\Psi_m(\vr)=e^{im\varphi}\begin{pmatrix}
	e^{-i\pi/4}\chi_\uparrow(r)\\
	e^{i\pi/4}\chi_\downarrow(r)e^{i\varphi}\\
	e^{-i\pi/4}\eta_\downarrow(r)e^{-il\varphi}\\
	e^{i\pi/4}\eta_\uparrow(r)e^{-i(l-1)\varphi}
\end{pmatrix}.
\] 
We define $\tilde{\Psi}_0=(\chi_\uparrow, \chi_\downarrow, \eta_\downarrow, \eta_\uparrow)^T$ for later convenience. 

Similar to the previous analysis, we first look for non-degenerate Majorana zero-energy state. The Majorana condition $\Xi \Psi\propto \Psi$ fixes the value of $m$ to be $\frac{l-1}{2}$ for odd $l$. For even $l$, there is no integer $m$ satisfying Majorana condition so no Majorana zero mode exists. The radial part of BdG equation then becomes
\begin{align}
	&\begin{pmatrix}
		\mathcal{H}_r & f(r)\\
		f(r) & -\sigma_y\mathcal{H}_r\sigma_y
	\end{pmatrix}\tilde{\Psi}_0(r)=0\\
	\mathcal{H}_r&=\begin{pmatrix}
		-\mu & v\left(\partial_r+\frac{m+1}{r}\right)\\
		-v\left(\partial_r-\frac{m}{r}\right) & -\mu
	\end{pmatrix}
	\label{eq:tiscradial}.
\end{align}
Here $\tilde{\Psi}_0$ is assumed real. Since we are interested in non-degenerate solution, $\Psi_0$ must be simultaneously an eigenstates of $\sigma_y\tau_y$ (particle-hole symmetry). This condition implies that $\eta_\uparrow=-\lambda \chi_\uparrow, \eta_\downarrow=\lambda \chi_\downarrow$ where $\lambda=\pm 1$. Taking into account above constraints $4 \times 4$ BdG equation reduces to 
\begin{equation}
	\begin{pmatrix}
		-\mu & v\left(\partial_r+\frac{m+1}{r}\right)+\lambda f\\
		-v\left(\partial_r-\frac{m}{r}\right)-\lambda f& -\mu
	\end{pmatrix}\begin{pmatrix}
		\chi_\uparrow\\
		\chi_\downarrow
	\end{pmatrix}=0. 
	\label{eq:radiatisc2}
\end{equation}
The solution of the above equation can be easily obtained for $\mu\neq 0$: 
\begin{equation}
	\begin{pmatrix}
		\chi_\uparrow\\
		\chi_\downarrow
	\end{pmatrix}=\mathcal{N}_3\begin{pmatrix}
		J_m(\frac{\mu}{v}r)\\
		J_{m+1}(\frac{\mu}{v}r)
	\end{pmatrix}e^{-\lambda\int_0^r \di r'\,f(r')}. 
	\label{eq:tiscsol}
\end{equation}
Obviously, we should take $\lambda=1$ to make radial wave functions normalizable. Here $\mathcal{N}_3$ is the normalization constant whose analytical form is given in \eqref{eq:normtisc}.

The case of $\mu=0$ is special due to the presence of an additional symmetry of BdG Hamiltonian - chiral symmetry. Imposing the boundary condition at $r\rightarrow 0$ that wave function must remain finite, for $l>0$ the solution of Eq.\eqref{eq:radiatisc2} becomes
\begin{equation}
	\begin{pmatrix}
		\chi_\uparrow\\
		\chi_\downarrow
	\end{pmatrix}\propto\begin{pmatrix}
		r^m\\
		0
	\end{pmatrix}e^{-\int_0^r \di r'\,f(r')}
	\label{eq:mu0zero1},
\end{equation}
and if $l<0$, $m+1<0$,
\begin{equation}
	\begin{pmatrix}
		\chi_\uparrow\\
		\chi_\downarrow
	\end{pmatrix}\propto\begin{pmatrix}
		0\\
		r^{-(m+1)}
	\end{pmatrix}e^{-\int_0^r \di r'\,f(r')}.
	\label{eq:mu0zero2}
\end{equation}
Particle-hole symmetry combined with the analyticity of the above solutions at $r \rightarrow 0$ constraints the integer $0 \leq m \leq l-1$. Similar conditions apply to the negative $l$ solutions. Thus, there are exactly $|l|$ zero-energy modes for any $l$ as found previously in Ref.~\cite{Jackiw_NPB81}. If $l$ is even, all these solutions are Dirac fermionic modes. However, if $l$ is odd, there are $l-1$ Dirac fermionic modes and one Majorana zero-energy mode as given in Eqs. \eqref{eq:mu0zero1} and \eqref{eq:mu0zero2}.
Because the chiral symmetry also relates eigenstates with positive energies to those with negative energies which follows from $\gamma^5\mathcal{H}_\text{BdG}\gamma^5=-\mathcal{H}_\text{BdG}$, one can always require the zero-energy eigenstates to be eigenstates of $\gamma^5$. The wave function in Eq.\eqref{eq:mu0zero1} is an eigenstate of $\gamma^5$ with eigenvalue $1$ while wave function \eqref{eq:mu0zero2} has eigenvalue $-1$. We define eigenstates of $\gamma^5$ with eigenvalue $\pm 1$ as $\pm$ chirality.

To summarize we have obtained the Majorana zero-energy bound state attached to the vortex with odd vorticity:
\begin{equation}
	\Psi_0(\vr)=e^{i(l-1)\varphi/2}\begin{pmatrix}
		e^{-i\pi/4}\chi_\uparrow(r)\\
		e^{i\pi/4}\chi_\downarrow(r)e^{i\varphi}\\
		e^{-i\pi/4}\chi_\downarrow(r)e^{-il\varphi}\\
	-e^{i\pi/4}\chi_\uparrow(r)e^{-i(l-1)\varphi}
\end{pmatrix}
	\label{eq:zeromodetisc}
\end{equation}
Generalization to the case of many vortices is straightforward. Order parameter with $2N$ vortices pinned at $\vec{R}_i$ is already given in \eqref{eq:mulvortices}. Assuming that they are well separated from each other, we can find an approximate zero-energy bound state localized in each vortex core:
\begin{equation}
	\Psi_i(\vr)\!=\!e^{i(l-1)\varphi_i/2}e^{i\Omega_i\tau_z/2}\begin{pmatrix}
		e^{-i\pi/4}\chi_\uparrow(r_i)\\
		e^{i\pi/4}\chi_\downarrow(r_i)e^{i\varphi_i}\\
		e^{-i\pi/4}\chi_\downarrow(r_i)e^{-il\varphi_i}\\
	-e^{i\pi/4}\chi_\uparrow(r_i)e^{-i(l-1)\varphi_i}
\end{pmatrix}
	\label{eq:zeromodetisc2}
\end{equation}
the construction of $N$ Dirac fermions and $2^{N-1}$ ground state Hilbert space are the same as the case of spinless $p_x+ip_y$ superconductors. 
\section{Degeneracy splitting due to intervortex tunneling}
\label{sec:splitting}
The ground state degeneracy, which is crucial for topological quantum computation with non-Abelian anyons, heavily relies on the assumption that intervortex tunneling is negligible. When tunneling effects are taken into account zero energy bound states are usually splitted and the ground state degeneracy is lifted. Besides, the sign of energy splitting is important for understanding many-body collective states~\cite{Ludwig_arxiv'10}. 

We now discuss a general formalism to calculate the energy splitting. We focus on the case of two classical vortices each with vorticity $l=1$ located at certain fixed positions $\vec{R}_1$ and $\vec{R}_2$. To develop a physical intuition, it is useful to view a vortex as a potential well, which may host bound states including zero-energy states, while the regions where superconducting gap is finite play the role of  a potential barrier. Therefore, the two-vortex problem  resembles the double-well potential problem in single-particle quantum mechanics (sometimes referred to as the Lifshitz problem in the literature~\cite{Landau_book3}). The solution to this simple problem in one-dimensional quantum mechanics is readily obtained~\cite{Landau_book3}  by considering symmetric and antisymmetric combinations of single-well wave-functions (which can be taken within the quasiclassical approximation for high barriers) and the overlap of these wave-functions always selects the symmetric state as the ground state in accordance with the elementary oscillation theorem ({\em i.e.}, the ground state has no nodes). We note that both  quasiclassical approximation and the Lifshitz method are not specific to the Schr{\"o}dinger equation, but actually represent general mathematical methods of solving differential equations of certain types. Moreover, these methods can be applied to rather generic matrix differential operators, and  such a generalization has been carried out by one of the authors in a completely different context of magnetohydrodynamics,~\cite{VGMHD} where interestingly the relevant differential operator appears to be mathematically similar to the BdG Hamiltonian. These considerations suggest that one can use the generalized Lifshitz method to obtain the splitting of zero modes of the BdG equations, by considering certain linear combinations of the individual Majorana modes in the two vortices and calculating their overlap, which reduces to a boundary integral along a path between the two vortices. Also, if the inter-vortex separation is large, one can use the semiclassical form of the Majorana wave-functions (effectively their large-distance asymptotes) to obtain quantitatively accurate results. Let us note here that apart from a technically more complicated calculation that needs to be carried out for the BdG equation, another important difference between this problem and the simple Lifshitz problem is that we can not rely on any oscillation theorem and there is no way to determine a priori which state has a lower energy. As discussed below, this ``uncertainty'' is fundamental to this problem and is eventually responsible for a fast-oscillating 
energy splitting with intervortex separation.

With the two zero-energy eigenstates $\Psi_1$ and $\Psi_2$ localized at $\vec{R}_1$ and $\vec{R}_2$ (given by Eq.\eqref{eq:zeromode2} for spinless $p_x+ip_y$ superconductor and by Eq.\eqref{eq:zeromodetisc} for TI/SC heterostructure), we can construct approximate eigenstate wave functions in the case of two vortices: $\Psi_\pm=\frac{1}{\sqrt{2}}(\Psi_1\pm e^{i\alpha}\Psi_2)$ analogous to the symmetric and anti-symmetric wave functions in a double-well problem with energies $E_\pm$, respectively. The phase factor $e^{i\alpha}$ can be determined from particle-hole symmetry which requires that new eigenstates $\Psi_+$ with energy $E_+=\delta E$ and $\Psi_-$ with energy $E_-=-\delta E$ be related by $\Xi\Psi_+=\Psi_-$. Since $\Psi_1$ and $\Psi_2$ are real (Majorana) eigenstates, one finds $\Xi\Psi_+=\frac{1}{\sqrt{2}}(\Psi_1+e^{-i\alpha}\Psi_2)=\Psi_-$. Thus, one arrives at $e^{2i\alpha}=-1$ which fixes $\alpha=\pm\pi/2$. In the rest of the text we take $\alpha=\pi/2$ for convenience. The corresponding quasiparticle operator can be identified with the Dirac fermion operator. We explicitly show this for the case of spinless $p_x+ip_y$ superconductor: 
\begin{equation}
  \hat{c}\!=\!\frac{1}{\sqrt{2}}(\hat{\gamma}_1\!-\!i\hat{\gamma}_2)\!=\!\int \di^2\vec{r}\,\left[\hat{\psi}\frac{u_1^\ast-\ii u_2^\ast}{\sqrt{2}}\!+\!\hat{\psi}^\dag\frac{v_1^\ast-\ii v_2^\ast}{\sqrt{2}}\right].
\end{equation}
Therefore $\hat{c}$ ($\hat{c}^\dag$) annihilates(creates) a quasiparticle on energy level $E_+$. The original two fold degeneracy between state with no occupation $\hat{c}|0\rangle=0$ and occupied $\ket{1}=\hat{c}^\dag\ket{0}$ is lifted by energy splitting $E_+$.

\newcommand{\rv}{{\mathrm{v}}}

To calculate the energy of $\Psi_+$, we employ the standard method based on the wave function overlap~\cite{Landau_book3}. Suppose the two vortices are placed symmetrically with respect to $y$ axis: $\vec{R}_1=(R/2,0)$ and $\vec{R}_2=(-R/2,0)$. BdG equations are $\mathcal{H}_\text{BdG}\Psi_+=E_+\Psi_+, \mathcal{H}_\text{BdG}\Psi_1=0$. We then multiply the first equation by $\Psi_1^*$ and second by $\Psi_+^*$, substract corresponding terms, and integrate over region $\Sigma$ which is the half plane $x\in (0,\infty), y\in(-\infty, \infty)$ arriving finally at the following expression for $E_+$:
\begin{align}\label{eq:overlap}
E_+=\frac{\int_\Sigma \di^2\vec{r}\,\Psi_1^\dag\mathcal{H}_{\mathrm{BdG}}\Psi_+-\int_\Sigma \di^2\vec{r}\,\Psi_+^\dag\mathcal{H}_{\mathrm{BdG}}\Psi_1}{\int_\Sigma\di^2\vec{r}\,\Psi_1^\dag\Psi_+}.
\end{align}
This is the general expression for the energy splitting which is used to evaluate $E_+$ in $p_x+ip_y$ SC and TI/SC heterostructure.\\

%%%%%%%%%%%%%%%%%%%%%%%%%%%%%%%%
\subsection{Splitting in spinless $p_x+ip_y$ superconductor}
%%%%%%%%%%%%%%%%%%%%%%%%%%%%%%%%

We now calculate splitting for two vortices in spinless $p_x+ip_y$ superconductor. The denominator in Eq.\eqref{eq:overlap}  can be evaluated quite straightforwardly $\int_\Sigma\di^2\vec{r}\,\Psi_1^\dag\Psi_+\approx 1/\sqrt{2}$. With the help of Green's theorem the integral over half plane in the numerator can be transformed into a line integral along the boundary of $\Sigma$, namely the $y$ axis at $x=0$ which we denote by $\partial\Sigma$:
\begin{align}
\begin{split}
%\int_\Sigma &\di^2\vec{r}\,  \Psi_1^\dag\mathcal{H}_{\mathrm{BdG}}  \Psi_+-\int_\Sigma \di^2\vec{r}\,\Psi_+^\dag\mathcal{H}_{\mathrm{BdG}}\Psi_1\\
%&=-\frac{i}{2m}\int_{\partial\Sigma}\left(u_1^\ast\partial_x u_2\!+\!u_2^\ast\partial_x u_1\right)\!+\!i\int_{\partial\Sigma}\Delta u_1^* u_2^*+\text{c.c}\\
E_+\!&=\!\frac{2}{m}\!\int_{\partial\Sigma}d y\!\left[g(s)g'(s)\cos 2\varphi_2\cos\varphi_2\! \right. \\
& \left. +\!\frac{g^2(s)}{s}\sin 2\varphi_2\sin\varphi_2-\frac{g^2(s)}{\xi}\right]
\end{split}\label{eq:splitting}
\end{align}
where $s=\sqrt{(R/2)^2+y^2},\,\tan\varphi_2=2y/R$. The function $g(s)$ is defined as $g(s)\equiv \chi(s)\exp(-s/\xi)$.

First we consider the regime where $\Delta_0^2<2m\mu v_F^2$ and radial wave function of Majorana bound state has the form \eqref{eq:chi1}. We are mainly interested in the behavior of energy splitting at large $R \gg \xi$ with $\xi$ being the coherence length, where our tunneling approximation is valid. Another length scales in our problem is the length corresponding to the bound state wave function oscillations $k=\sqrt{2m\mu-\Delta_0^2/v_F^2}$. In the limit $R\gg \text{max}(k^{-1},\xi)$ upon evaluating the integral~\eqref{eq:splitting} we obtain
\begin{widetext}
\begin{equation}
  E_+=\sqrt{\frac{8}{\pi}}\frac{\mathcal{N}_1^2}{m}\left(\frac{\lambda^2}{1+\lambda^2}\right)^{1/4}\!\frac{1}{\sqrt{kR}}\exp\left(-\!\frac{R}{\xi}\right)\left[\cos(kR+\alpha)
-\frac{2}{\lambda}\sin(kR+\alpha)+\frac{2(1+\lambda^2)^{1/4}}{\lambda}\right],
\label{eq:splittinggeneral}
\end{equation}
\end{widetext}
where $\lambda=k\xi,\, 2\alpha=\arctan \lambda$ and $\mathcal{N}_1$ is the normalization constant defined in Eq.\eqref{eq:chi1}. The expression of $\mathcal{N}_1$ is given in Appendix \ref{sec:norm} and has the following asymptotes for $\lambda\gg 1$ and $\lambda\ll 1$:
\begin{equation}
	\mathcal{N}_1^2=\begin{cases}
		\frac{k}{2\xi} & \lambda\gg 1\\
		\frac{8}{3\pi k^2\xi^4} & \lambda \ll 1
	\end{cases}
	\label{eq:norm1asymp}.
\end{equation}
 The exponential decay is expected due to the fact that Majorana bound states are localized in vortex core. In addition, the splitting energy $E_+$ oscillates with intervortex seperation $R$ which can be traced back to interference between the wave functions of the two Majorana bound states since they both oscillates in space. 

Of particular importance is the sign of splitting as noted in Ref.~\cite{Ludwig_arxiv'10}. It determines which state is energetically favored when tunneling interaction is present. If $E_+>0$, $\ket{0}$ is favored whereas $E_+<0$ favors $\ket{1}$.  We note here that the definition of states $\ket{0}$ and $\ket{1}$ relies on how we define the Dirac fermion operator $\hat{c}$ and $\hat{c}^\dag$. Due to the presence of a constant term together with trigonometric function, the sign of splitting can change. To figure out when the sign oscillates, we require the amplitude of the trigonometric part is greater than the constant part which gives
\[
\sqrt{1+\frac{4}{\lambda^2}}>\frac{2(1+\lambda^2)^{1/4}}{\lambda}.
\]
Solving this inequality yields $\lambda=k\xi>8$. Therefore in this parameter regime the sign of splitting changes with distance $R$. Otherwise the splitting still shows oscillatory behavior but the sign is fixed to be positive. 

In weak-coupling superconductors where $\Delta_0\ll \varepsilon_F$ or equivalently $k_F\xi\gg 1$, the expression for the energy splitting \eqref{eq:splittinggeneral} can be considerably simplified. In this case, $\mu\approx \varepsilon_F$ and $k\approx k_F$. Keeping only terms that are leading order in $(k_F\xi)^{-1}$, we find
%Use the explicit expression for $\chi(r)$ given in \eqref{eq:chi1} we have
%\begin{equation}
%E_+ \approx - \frac{\sqrt{2}\Delta_0 a}{\pi}\!\int_{-\infty}^{\infty}\!\di y\!\, \frac{\cos \left(2\lambda s\right)}{s^2}\,\!\exp\!\left(\!-2s\right)\nonumber\!,
%\end{equation}
%where $\lambda=k_F\xi, a=R/2\xi$. Assuming $\lambda\gg 1$ and $a\gg 1$, this integral can be evaluated to give,
\begin{equation}
  E_+\approx \sqrt{\frac{2}{\pi}}\Delta_0\frac{\cos(k_FR+\frac{\pi}{4})}{\sqrt{k_FR}}\exp\left(-\frac{R}{\xi}\right),
  \label{eq:splitting1}
\end{equation}
which is the expression reported in Ref.~[\onlinecite{Cheng_PRL09}]. A similar expression for splitting of a pair of Majorana bound states on superconductor/2D topological insulator/magnet interface is found in Ref.~[\onlinecite{Nilsson_PRL08}].

Next we consider a different limit $\Delta_0^2>2m\mu v_F^2$ in which the wave function of Majorana bound state for a single vortex doesn't show any spatial oscillations.  Thus, we expect that tunneling splitting will show just an exponential decay without any oscillations. The wave function \eqref{eq:chi2} grows exponentially when $r\rightarrow \infty$:
\[
\chi(r)\sim \frac{1}{\sqrt{r}}e^{k_0r}
\]
with $k_0=\sqrt{\Delta_0^2/v_F^2-2m\mu}$. The overall radial wave function decays exponentially $\sim \exp(-k'r)$ where $k'=1/\xi-k_0$. In this case, the tunneling approximation is only valid for $k'R\gg 1$ since bound state wave function is localized approximately within distance $1/k'$ to vortex core. The resulting energy splitting monotonically decays:
\begin{equation}
  E_+\approx \sqrt{\frac{2}{\pi}}\frac{\mathcal{N}_2^2}{m}\left(\frac{3}{k_0\xi}-1\right)\frac{1}{\sqrt{k'R}}\exp(-k'R),
	\label{eqn:splitting3}
\end{equation}
where the normalization $\mathcal{N}_2$ is defined in Eq.\eqref{eq:nc2}. As $\mu$ approaches $0$ there is a quantum phase transition between the non-Abelian phase and Abelian phase. This transition is accompanied by closing of the gap and the Majorana bound state is no longer localized since $k'\rightarrow 0$.

We briefly comment on the degeneracy splitting between vortex zero modes in the ferromagnetic insulator/semiconductor/superconductor hybrid structure proposed by Sau {\it et. al.}~\cite{Sau_PRL10} which can be modeled by spin-$1/2$ fermions with Rashba spin-orbit coupling and s-wave pairing induced by the superconducting proximity effect. Since time-reversal symmetry is broken by the proximity-induced exchange splitting, this system belongs to the same symmetry class as spinless $p_x+ip_y$ superconductor - class D. The connection between this hybrid structure and spinless $p_x+ip_y$ can be made more explicit by the following argument: the single particle Hamiltonian after diagonalization yields two bands. Assuming a large band gap (which is actually determined by exchange field), one can project the full Hamiltonian onto the lower band and then the effective Hamiltonian  takes exactly the form of spinless $p_x+ip_y$ superconductor, see, for example, the discussion in Ref. [\onlinecite{Alicea_PRB10}]. Although analytical expression for Majorana bound state in vortex core is not available, the solution behaves qualitatively similar to the one in spinless $p_x+ip_y$ superconductor. Therefore, we expect that splitting should also resemble that of spinless $p_x+ip_y$ superconductor.

%%%%%%%%%%%%%%%%%%%%%%%
\subsection{Splitting in TI/SC heterostructure}
%%%%%%%%%%%%%%%%%%%%%%%

In this section we discuss the case of vortex-vortex pair in TI/SC heterostructure. We assume both vortices have vorticity $1$. Similar the case of  $p_x+ip_y$ superconductor, one can transform the surface integral over half plane $\Sigma$ to a line integral along its boundary $\partial\Sigma$. Exploiting the explicit expressions for the zero mode solution, we arrive at
\begin{equation}
\begin{split}
\int_\Sigma \di^2\vec{r}&\,  \Psi_1^\dag\mathcal{H}_{\mathrm{BdG}}  \Psi_+-\int_\Sigma \di^2\vec{r}\,\Psi_+^\dag\mathcal{H}_{\mathrm{BdG}}\Psi_1\\
&=-2\sqrt{2}v\int_{-\infty}^\infty \di y\,\chi_\uparrow(s)\chi_\downarrow(s)\cos\varphi_2,
\label{eq:splittingtisctmp}
\end{split}
\end{equation}
where $s=\sqrt{(R/2)^2+y^2},\,\cos\varphi_2=R/2s$. 

First we consider the case with finite $\mu$. There are two length scales: Fermi wavelength $k_F^{-1}=\frac{v}{\mu}$ and coherence length $\xi=\frac{v}{\Delta_0}$. We evaluate the integral \eqref{eq:splittingtisctmp} in the limit where $R$ is large compared to both $k_F^{-1}=v/\mu$ and $\xi$:
\begin{equation}
  E_+\!\approx\!\frac{4\mathcal{N}_3^2v}{\sqrt{\pi}k_F(1+k_F^2\xi^2)^{1/4}}\frac{\cos(k_FR+\alpha)}{\sqrt{R/\xi}}\exp\left(\!-\frac{R}{\xi}\right),
	\label{eq:splittingtisc}
\end{equation}
where $2\alpha=\arctan(k_F\xi)$ and the normalization $N_3$ is given by Eq.\eqref{eq:normtisc}. One can notice that the splitting, including its sign, oscillates with the intervortex separation $R$ when $R$ is large. In the limit of large $\mu$, say $k_F\xi\gg 1$, Eq. \eqref{eq:splittingtisc} can be simplified to
\begin{equation}
  E_+\approx \frac{2\Delta_0}{\sqrt{\pi}}\frac{\cos(k_FR+\frac{\pi}{4})}{\sqrt{k_FR}}\exp\left(-\frac{R}{\xi}\right).
  \label{eq:splittingtisc1}
\end{equation}

We now turn to the limit where $\mu$ is very close to Dirac point, {\it i.e.} $\mu\rightarrow 0,\,k_F\xi\ll 1$. We evaluate the integral for $\xi \ll R\ll k_F^{-1}$. \begin{equation}
  E_+ \approx -\frac{2\mu}{\sqrt{\pi}}\left(\frac{R}{\xi}\right)^{3/2}\exp\left(-\frac{R}{\xi}\right),
	\label{eq:splittingdirac2}
\end{equation}
where we have made use of asymptote of $\mathcal{N}_3$ in the limit $k_F\xi\ll 1$. Eq. \eqref{eq:splittingdirac2} implies that for fixed $R$ the splitting vanishes as $\mu$ approaches Dirac point. Actually this fact can be easily seen from \eqref{eq:splittingtisctmp} without calculating the integral. Because at $\mu=0$ either $\chi_\uparrow$ or $\chi_\downarrow$ vanishes, the splitting which is proportional to the product of $\chi_\downarrow$ and $\chi_\uparrow$ is zero. The same result for splitting at $\mu=0$ has also been obtained in Ref.~[\onlinecite{Chamon_PRB10}]. 

We now show that vanishing of the splitting at $\mu=0$ is a direct consequence of chiral symmetry.  At $\mu=0$ zero modes carry chirality which labels the eigenvalues of $\gamma^5$. More specifically, wave function is an eigenstate of $\gamma^5$: $\gamma^5\Psi_i=\lambda\Psi_i$. Consider an arbitrary perturbation represented by $\mathcal{O}$ to the ground state manifold expanded by these local zero modes. To leading order in perturbation theory its effect is determined by matrix element $\mathcal{O}_{ij}=\langle \Psi_i|\mathcal{O}|\Psi_j\rangle$. Now assume that $\Psi_i$ and $\Psi_j$ have the same chirality(which means that vortices $i$ and $j$ have identical vorticity). If the perturbation $\mathcal{O}$ preserves chiral symmetry, {\it i.e.} $\{\gamma^5,\mathcal{O}\}=0$, then
\begin{equation}
  \langle\Psi_i|\{\gamma^5,\mathcal{O}\}|\Psi_j\rangle=2\lambda\langle \Psi_i|\mathcal{O}|\Psi_j\rangle=0.
  \label{eq:matrixelementiszero}
\end{equation}
Therefore matrix element $\langle\Psi_i|\mathcal{O}|\Psi_j\rangle$ vanishes identically. Tunneling obviously preserves chiral symmetry so there is no splitting between two vortices with the same vorticity from this line of reasoning. As discussed below this fact actually holds beyond perturbation theory and the robustness of zero modes in the presence of chiral symmetry is ensured by an index theorem.

\section{Atiyah-Singer-type index theorem}\label{sec:index}
Index theorem provides an intelligent way of understanding the topological stability of zero modes. It is well-known that one can relate the analytical index of an elliptic differential operator (Dirac operator) %(which is equal to the number of zero modes) 
to the topological index (winding number) of the background scalar field in 2D~[\onlinecite{Weinberg_PRD81}]) through the index theorem. Since BdG Hamiltonian for TI/SC system at $\mu=0$ can be presented as a Dirac operator (see Eq.\eqref{eq:Dirac}), we give a brief review of this index theorem, see also recent exposition in Ref.~[\onlinecite{Fukui_JPSJ10}]. Specifically, the Hamiltonian for TI/SC heterostructure can be written as
\begin{equation}
	\mathcal{H}_D=i \bm \gamma \cdot \bm \nabla+\bm \Gamma \cdot \bm n, 
	\label{eq:diracop}
\end{equation}
where $\bm n =(\Re \Delta, -\Im \Delta)$ field describes the non-trivial configuration of the superconducting order parameter. We assume the following boundary condition for $\bm n$ field: 
\begin{equation}
	|\vec{n}(\vr)|\rightarrow \text{const} \:\text{as}\: |\vec{r}|\rightarrow \infty
	\label{eq:bcphi}.
\end{equation}
As  mentioned above, this model Hamiltonian has particle-hole symmetry, time-reversal symmetry and chiral symmetry which is given by $\gamma^5$. It anticommutes with the Dirac Hamiltonian $\{\gamma^5,\mathcal{H}_D\}=0$. Therefore, all zero modes $\Psi_0$ of $\mathcal{H}_D$ are eigenstates of $\gamma^5$. Since $(\gamma^5)^2=1$ eigenvalues of $\gamma^5$ are $\pm 1$. We define $\pm$ chirality of zero modes as $\gamma^5\Psi_0^\pm=\pm\Psi_0^\pm$. The analytical index of $\mathcal{H}_D$ is defined as
\begin{equation}
	\text{ind}\, \mathcal{H}_D=n_+-n_-,
	\label{eq:ind1}
\end{equation}
where $n_\pm$ are number of zero modes with $\pm$ chirality.

The index theorem for the Hamiltonian $H_D$ states that the analytical index is identical to the winding number of the background scalar field in the two-dimensional space~\cite{Weinberg_PRD81}:
\begin{equation}
	\text{ind} \,\mathcal{H}_D=-\frac{1}{2\pi}\int\di^ix \,\epsilon^{ab}\hat{ n}_a\partial_i\hat{n}_b,
	\label{eq:indextheorem}
\end{equation}
%Although we here restrict ourselves to the case $d=2$ the theorem can be generalized to higher dimensional space. 
where $\hat{\bm n}=\vec{\bm n}/|\bm n|$. According to the index theorem, the number of zero modes is determined by the topology of order parameter at infinity. The right hand side is ensured to be an integer by the fact that the homotopy group $\pi_{1}(S^{1})=\mathbb{Z}$. If we have a vortex in the system with vorticity $l$, the right hand side of \eqref{eq:indextheorem} evaluates exactly to $l$. Thus the index theorem implies that the Dirac Hamiltonian has at least $l$ zero modes which agrees with explicit solution obtained by Jackiw and Rossi~\cite{Jackiw_NPB81}. This conclusion can be generalized to the case where multiple vortices are present. In that case the right hand side is basically the sum of vorticities of all vortices.

The index theorem \eqref{eq:indextheorem} requires chiral symmetry which is broken by presence of a finite chemical potential $\mu\neq 0$. Now we argue that when chiral symmetry is broken the Majorana zero modes admit a $\mathbb{Z}_2$ classification corresponding to even-odd number of zero energy solutions. Generally speaking, a small chiral symmetry breaking term cause coupling between zero modes and split them away from zero energy. However, due to particle-hole symmetry, the number of zero modes that are split by chiral symmetry breaking term must be even. So the parity of the topological index is preserved in the generic case. This is consistent with an explicit solutions of zero mode in TI/SC heterostruture with finite chemical potential. Thus, we conclude that without chiral symmetry the Majorana zero modes bound to vortices are classified by $\mathbb{Z}_2$ corresponding to even or odd number of zero modes.

%Although a mathematically rigorous proof is not available, we can confidently conclude that in the generic case without chiral symmetry, the Majorana zero modes bound to vortices are classified by $\mathbb{Z}_2$.

Now we can fit our splitting calculation into the general picture set by index theorem. As being argued above, Majorana zero modes in spinless $p_x+ip_y$ superconductor is classified by $\mathbb{Z}_2$. When there are two vortices in the bulk, the topological index of order parameter is $2$ thus there is no zero mode and we find the splitting as expected. The same applies to two vortices in TI/SC heterostructure with $\mu\neq 0$. However, as we have seen in the calculation the splitting vanishes for $\mu=0$. This should not be surprising since according to index theorem, there should be at least two zero modes associated with total vorticity $2$ which is the case for two vortices.

\subsection{Comparison with the splitting calculations in other systems.}
Recently numerical calculations of the degeneracy splitting have been performed for other systems supporting non-Abelian Ising anyons~\cite{Tserkovnyak_PRL03, Baraban_PRL09, Lahtinen_AnnPhys08}. 
In all these calculations it was found that the splitting has qualitatively similar behavior - there is an exponential decay with the oscillating prefactor which stems from the spatial oscillations of Majorana bound states. In the case of  Moore-Read quantum Hall state~\cite{Baraban_PRL09}, the splitting between two quasiholes exponentially decays and oscillates with the magnetic length $l_c=\frac{\hbar }{e B}$ since there only one length scale in the problem. The oscillatory behavior is also predicted for pair of vortex excitations in the B-phase of Kitaev's honeycomb lattice model in an external magnetic field~\cite{Lahtinen_AnnPhys08}.

\section{Collective states of many-anyon system}
\label{sec:collective}

The microscopic calculations of the degeneracy splitting for a pair of vortices are important for understanding the collective states of anyons arising on top of the non-Abelian parent state when many Majorana fermions (Ising anyons) are present~\cite{Grosfeld_prb'06, Feiguin_prl'07, Gils_prl'09, Lahtinen_PRB'10, Ludwig_arxiv'10}. Essentially, the sign of the splitting favors certain fusion channel ( $\bm 1$ or $\psi$ in the terminology of Ref.[\onlinecite{nayak_RevModPhys'08}]) when two vortices carrying Majorana fermions are brought together. These fusion channels correspond to having a fermion ($\psi$-channel when $E_+<0$) or no fermion  ($\bm 1$-channel when $E_+>0$) left upon fusing of two anyons. 

For pedagogical reason we start with the dilute anyon density limit assuming that the average distance between Majorana fermions is large compared with the coherence length $\xi$. In this regime, the many anyon state of the system will resemble gas of weakly bound pairs of anyons formed out of two anyons separated by the smallest distance. Because of the exponential dependence of the energy splitting the residual ``interactions" with other anyons are exponentially smaller and can be ignored. In this scenario the parent state remains unchanged.  

%In the Quantum Hall context qualitatively similar scenario was studied by Bonderson and
%Slingerland~\cite{Bonderson_prb'08} who argued that charge $e/4$ quasiholes can pair up to form $e/2$ Abelian molecules which in turn form non-Abelian hierarchical states. %The condensation of quasiparticles and non-Abelian daughter states emerging on top of Moore-Read state was recently discussed in Ref.~[\onlinecite{Hermanns_prl'10}] again yielding the non-Abelian daughter state. 
%However, the analogy between Majorana fermions in the chiral superconducting and Fractional Qauntum Hall $\nu=5/2$ states is not rigorous, and should be regarded at the qualitative level (e.g., Majorana fermions in superconductors are neutral whereas in QH states they are charged).   

When the density of anyons is increased so that the average distance between them becomes of the order of the Majorana bound state decay length (coherence length $\xi$ in p-wave superconductors or magnetic length $l_c$ in Quantum Hall states) the system can form a non-trivial collective liquid  (Wigner crystal of anyons or some other incompressible liquid state). 
This question has been investigated in Refs.~\cite{Bonderson_prb'08, Levin_prb'09, Hermanns_prl'10, Ludwig_arxiv'10}. Although our approach used to calculate energy splitting breaks down in this regime and one should resort to numerical calculations for the magnitude of the splitting, we believe that qualitative form of the splitting will remain the same. It is interesting to discuss the collective state that forms in this regime.   Remarkably, it was shown in Ref.~[\onlinecite{Ludwig_arxiv'10}] that depending on the fusion channel ({\it i.e.} sign of the splitting) the collective state of anyons may be Abelian or non-Abelian.  This result was obtained assuming that the magnitude of the splitting is constant and the sign of the splitting is the same for all anyons (positive or negative). However, because of the prefactor changing rapidly with the Fermi wave length we expect the magnitude of the splitting energy to be random realizing random bond Ising model discussed in Ref.~[\onlinecite{read_prb'00_1, Ludwig_arxiv'10}].     

Finally, we mention that our calculations above and all existing studies of interacting many-anyon systems treat host vortices as classical objects with no internal dynamics. This is a well-defined mathematical framework, which corresponds to the BCS mean-field approximation. In real superconductors, however, there are certainly corrections to it. The order parameter field, $\Delta({\bf r},t)$, which describes a certain vortex configuration has  a non-trivial dynamics and fluctuates in both space and time. At low temperatures, when the system is fully gapped, these fluctuation effects are suppressed in the bulk, but they are always significant in the vicinity of the vortex core, where the order parameter vanishes. This dynamics gives rise to an effective motion of a vortex as well as to the dynamics of its shape and the radial profile. The relevant length-scales of these effects certainly exceed the Fermi wave-length, which is the smallest length-scale in the problem in most realistic systems. Even if the vortex is pinned, e.g.  by disorder, its motion can be constrained only up to a mean-free path or another relevant length-scale, which is still much larger than the Fermi wave-length for local superconductivity to exist. These considerations suggest that the intervortex separation between quantum vortices has an intrinsic quantum uncertainty, which is expected to much exceed the inverse Fermi wave-vector. This makes the question of the sign of Majorana mode coupling somewhat ill-defined in the fully quantum problem. Indeed we found the energy splitting to behave as $\delta E({\bf r}) = \left| \delta E_0(r) \right| \cos \left( k_F r+\alpha \right)$, where $\left| \delta E_0(r) \right|$
is an exponentially small magnitude of coupling insensitive to any dynamics of ${\bf r}(t)$. The cosine-factor, which determines the sign, is however expected to be very much sensitive to quantum dynamics. To derive the actual microscopic model even in the simplest case of two non-Abelian anyons living in the cores of quantum vortices is a tremendously complicated problem, which requires a self-consistent treatment of the vortex order-parameter field and fermionic excitations beyond mean-field. However, one can argue that the outcome of such a treatment would be an effective theory  where the $e^{i k_F { r}(t)}$ factor that appears in Majorana interactions, should be replaced with a random quantum-fluctuating phase ({\em c.f.}, Ref.~[\onlinecite{ZSVGAI}]), $e^{i \theta(t)}$, whose dynamics is governed by an effective action of type, $S[\theta] \approx \int d\tau \left[ \left(\theta - \theta_0\right)^2 + c \left(\partial_\tau \theta\right)^2 \right]$. This generally resembles a gauge theory, but of an unusual type, and at this stage it is unclear what collective many-anyon state such a theory may give rise to. 

\section{Conclusions}

In this paper, we address the problem of topological degeneracy lifting in topological superconductors characterized by the presence of Majorana zero-energy states bound to the vortex cores. We calculate analytically energy splitting of zero-energy modes due to the intervortex tunneling. We consider here canonical model of topological superconductor, spinless $p_x+ip_y$ superconductor, as well as the model of Dirac fermions coupled to superconducting scalar field. The latter is realized at the topological insulator/s-wave superconductor interface.     
%In low energy regime tunneling splitting is the dominant interaction present between non-Abelian anyons. The topological protection of ground state degeneracy is spoiled by a finite splitting of zero modes. We perform analytical calculation of tunneling splitting as a function of intervortex distance in the canonical example of non-Abelian topological superconducotr, the spinless $p_x+ip_y$ superconductor and also the model of Dirac fermions coupled to superconducting scalar field. 
In the case of spinless $p_x+ip_y$ superconductor, we find that, in addition to the expected exponential decay,  the splitting energy for a pair of vortices oscillates with distance in weak-coupling superconductor and these oscillations become over-damped as the magnitude of the chemical potential is decreased. In the second model, the splitting energy oscillates for finite chemical potential and vanishes at $\mu=0$.  The vanishing of splitting energy is a consequence of an additional symmetry, the chiral symmetry, emerging in the model when chemical potential is exactly equal to zero. We show that this fact is not accidental but stems from the index theorem which relates the number of zero modes of the Dirac operator to the topological index of the order parameter. Finally, we discuss the implications of our results for many-anyon systems. \\

\section{Acknowledgements}

We thank P. Bonderson, A. W. W. Ludwig, C. Nayak and S. Trebst for illuminating discussions. We acknowledge a stimulating correspondence with R. Jackiw regarding the index theorem. This work was supported by DARPA-QuEST. 

{\em Note added:~}After this manuscript was finalized for submission, we became aware of a related preprint~\cite{Mizushima_arxiv10_5} by T.~Mizushima and K.~Machida, which has some overlap with our results.\\
\appendix
\section{Normalization of Majorana bound state wave function}
\label{sec:norm}
In this appendix we present expressions for the normalization constants of Majorana bound state radial wave functions in Eqs.\eqref{eq:chi1},\eqref{eq:chi2} and \eqref{eq:tiscsol}. These constants are expressed in terms of hypergeometric functions.

Normalization constant $\mathcal{N}_1$ appearing in Eq.\eqref{eq:chi1} is defined as 
\begin{equation}
  4\pi\mathcal{N}_1^2\int_0^\infty r\di r\,J_1^2(k_1r)e^{-2r/\xi}=1,
  \label{eq:nc1formula}
\end{equation}
where $k_1=\sqrt{2m\mu-\Delta_0^2/v_F^2}$. Evaluation of the integral yields:
\begin{equation}
\mathcal{N}_1^2=\frac{8}{3\pi k_1^2\xi^4\,\, _2F_1(\frac{3}{2},\frac{5}{2}; 3; -k_1^2\xi^2)}
\label{eq:nc1}
\end{equation}
with its asymptotes given by
\begin{equation}
  \mathcal{N}_1^2\sim\begin{cases}
    \frac{8}{3\pi k_1^2\xi^4} & k_1\xi\ll 1\\
    \frac{k_1}{2\xi} & k_1\xi\gg 1
  \end{cases}.
  \label{eq:asympn1}
\end{equation}

Now we turn to $\mathcal{N}_2$. Similarly, it's determined by
\begin{equation}
  4\pi\mathcal{N}_2^2\int_0^\infty r\di r\,I_1^2(k_2r)e^{-2r/\xi}=1,
  \label{eq:nc2}
\end{equation}
where $k_2=\sqrt{\Delta_0^2/v_F^2-2m\mu}$. Since $\mu>0$, $k_2\xi$ is always smaller than $1$. We find $\mathcal{N}_2$ is given by
\begin{equation}
\mathcal{N}_2^2=\frac{8}{3\pi k_2^2\xi^4\,\, _2F_1(\frac{3}{2},\frac{5}{2}; 3; k_2^2\xi^2)}
\label{eq:nc2result}
\end{equation}
Finally, the normalization constant of wave function in \eqref{eq:tiscsol} can be calculated from
\begin{equation}
  4\pi\mathcal{N}_3^2\int_0^\infty r\di r\,\left[J_m^2(\frac{\mu}{v}r)+J_{m+1}^2(\frac{\mu}{v}r)\right]e^{-2r/\xi}=1,
  \label{eq:nc3}
\end{equation}
which yields
\begin{equation}
	\mathcal{N}_3^2=\frac{8}{\pi \xi^2\!\left[8\, {_2F_1}(\frac{1}{2},\frac{3}{2}; 1; -\lambda^2)\!+\!3\lambda^2\, {_2F_1}(\frac{3}{2},\frac{5}{2}; 3; -\lambda^2)\right]}
	\label{eq:normtisc}
\end{equation}
with $\lambda=\mu\xi/v$. It has the following asymptotes:
\begin{equation}
  \mathcal{N}_3^2\sim\begin{cases}
    \frac{1}{\pi\xi^2} & \lambda\ll 1\\
    \frac{\lambda}{2\xi^2} & \lambda\gg 1.
  \end{cases}
  \label{eq:asympn3}
\end{equation}
%\bibliography{./refs_pwave}

\begin{thebibliography}{82}
\expandafter\ifx\csname natexlab\endcsname\relax\def\natexlab#1{#1}\fi
\expandafter\ifx\csname bibnamefont\endcsname\relax
  \def\bibnamefont#1{#1}\fi
\expandafter\ifx\csname bibfnamefont\endcsname\relax
  \def\bibfnamefont#1{#1}\fi
\expandafter\ifx\csname citenamefont\endcsname\relax
  \def\citenamefont#1{#1}\fi
\expandafter\ifx\csname url\endcsname\relax
  \def\url#1{\texttt{#1}}\fi
\expandafter\ifx\csname urlprefix\endcsname\relax\def\urlprefix{URL }\fi
\providecommand{\bibinfo}[2]{#2}
\providecommand{\eprint}[2][]{\url{#2}}

\bibitem[{\citenamefont{Nayak et~al.}(2008)\citenamefont{Nayak, Simon, Stern,
  Freedman, and Das~Sarma}}]{nayak_RevModPhys'08}
\bibinfo{author}{\bibfnamefont{C.}~\bibnamefont{Nayak}},
  \bibinfo{author}{\bibfnamefont{S.~H.} \bibnamefont{Simon}},
  \bibinfo{author}{\bibfnamefont{A.}~\bibnamefont{Stern}},
  \bibinfo{author}{\bibfnamefont{M.}~\bibnamefont{Freedman}}, \bibnamefont{and}
  \bibinfo{author}{\bibfnamefont{S.}~\bibnamefont{Das~Sarma}},
  \bibinfo{journal}{Rev. Mod. Phys.} \textbf{\bibinfo{volume}{80}},
  \bibinfo{pages}{1083} (\bibinfo{year}{2008}).

\bibitem[{\citenamefont{Moore and Read}(1991)}]{Moore_NPB91}
\bibinfo{author}{\bibfnamefont{G.}~\bibnamefont{Moore}} \bibnamefont{and}
  \bibinfo{author}{\bibfnamefont{N.}~\bibnamefont{Read}},
  \bibinfo{journal}{Nucl. Phys. B} \textbf{\bibinfo{volume}{360}},
  \bibinfo{pages}{362 } (\bibinfo{year}{1991}).

\bibitem[{\citenamefont{Nayak and Wilczek}(1996)}]{Nayak_NPB96}
\bibinfo{author}{\bibfnamefont{C.}~\bibnamefont{Nayak}} \bibnamefont{and}
  \bibinfo{author}{\bibfnamefont{F.}~\bibnamefont{Wilczek}},
  \bibinfo{journal}{Nucl. Phys. B} \textbf{\bibinfo{volume}{479}},
  \bibinfo{pages}{529 } (\bibinfo{year}{1996}).

\bibitem[{\citenamefont{Greiter et~al.}(1992)\citenamefont{Greiter, Wen, and
  Wilczek}}]{Greiter_NPB92}
\bibinfo{author}{\bibfnamefont{M.}~\bibnamefont{Greiter}},
  \bibinfo{author}{\bibfnamefont{X.~G.} \bibnamefont{Wen}}, \bibnamefont{and}
  \bibinfo{author}{\bibfnamefont{F.}~\bibnamefont{Wilczek}},
  \bibinfo{journal}{Nucl. Phys. B} \textbf{\bibinfo{volume}{374}},
  \bibinfo{pages}{567 } (\bibinfo{year}{1992}).

\bibitem[{\citenamefont{Read and Green}(2000)}]{read_prb'00}
\bibinfo{author}{\bibfnamefont{N.}~\bibnamefont{Read}} \bibnamefont{and}
  \bibinfo{author}{\bibfnamefont{D.}~\bibnamefont{Green}},
  \bibinfo{journal}{Phys. Rev. B} \textbf{\bibinfo{volume}{61}},
  \bibinfo{pages}{10267} (\bibinfo{year}{2000}).

\bibitem[{\citenamefont{Volovik}(2003)}]{Volovik_book}
\bibinfo{author}{\bibfnamefont{G.~E.} \bibnamefont{Volovik}},
  \emph{\bibinfo{title}{The Universe in a Helium Droplet}}
  (\bibinfo{publisher}{Oxford University Express}, \bibinfo{year}{2003}).

\bibitem[{\citenamefont{Volovik}(1999)}]{Volovik_JETP'99}
\bibinfo{author}{\bibfnamefont{G.}~\bibnamefont{Volovik}},
  \bibinfo{journal}{JETP Lett.} \textbf{\bibinfo{volume}{70}},
  \bibinfo{pages}{609} (\bibinfo{year}{1999}).

\bibitem[{\citenamefont{Ivanov}(2001)}]{Ivanov_PRL'01}
\bibinfo{author}{\bibfnamefont{D.~A.} \bibnamefont{Ivanov}},
  \bibinfo{journal}{Phys. Rev. Lett.} \textbf{\bibinfo{volume}{86}},
  \bibinfo{pages}{268} (\bibinfo{year}{2001}).

\bibitem[{\citenamefont{Kitaev}(2003)}]{Kitaev_AP03}
\bibinfo{author}{\bibfnamefont{A.~Y.} \bibnamefont{Kitaev}},
  \bibinfo{journal}{Ann. Phys.(N.Y.)} \textbf{\bibinfo{volume}{303}},
  \bibinfo{pages}{2 } (\bibinfo{year}{2003}).

\bibitem[{\citenamefont{Das~Sarma et~al.}(2005)\citenamefont{Das~Sarma,
  Freedman, and Nayak}}]{dassarma_prl'05}
\bibinfo{author}{\bibfnamefont{S.}~\bibnamefont{Das~Sarma}},
  \bibinfo{author}{\bibfnamefont{M.}~\bibnamefont{Freedman}}, \bibnamefont{and}
  \bibinfo{author}{\bibfnamefont{C.}~\bibnamefont{Nayak}},
  \bibinfo{journal}{Phys.\ Rev.\ Lett.} \textbf{\bibinfo{volume}{94}},
  \bibinfo{pages}{166802} (\bibinfo{year}{2005}).

\bibitem[{\citenamefont{Tewari et~al.}(2007{\natexlab{a}})\citenamefont{Tewari,
  Das~Sarma, Nayak, Zhang, and Zoller}}]{tewari_prl'2007}
\bibinfo{author}{\bibfnamefont{S.}~\bibnamefont{Tewari}},
  \bibinfo{author}{\bibfnamefont{S.} \bibnamefont{Das~Sarma}},
  \bibinfo{author}{\bibfnamefont{C.}~\bibnamefont{Nayak}},
  \bibinfo{author}{\bibfnamefont{C.}~\bibnamefont{Zhang}}, \bibnamefont{and}
  \bibinfo{author}{\bibfnamefont{P.}~\bibnamefont{Zoller}},
  \bibinfo{journal}{Phys.\ Rev.\ Lett.} \textbf{\bibinfo{volume}{98}},
  \bibinfo{pages}{010506} (\bibinfo{year}{2007}{\natexlab{a}}).

\bibitem[{\citenamefont{Radu et~al.}(2008)\citenamefont{Radu, Miller, Marcus,
  Kastner, Pfeiffer, and West}}]{Radu_Sci08}
\bibinfo{author}{\bibfnamefont{I.~P.} \bibnamefont{Radu}},
  \bibinfo{author}{\bibfnamefont{J.~B.} \bibnamefont{Miller}},
  \bibinfo{author}{\bibfnamefont{C.~M.} \bibnamefont{Marcus}},
  \bibinfo{author}{\bibfnamefont{M.~A.} \bibnamefont{Kastner}},
  \bibinfo{author}{\bibfnamefont{L.~N.} \bibnamefont{Pfeiffer}},
  \bibnamefont{and} \bibinfo{author}{\bibfnamefont{K.~W.} \bibnamefont{West}},
  \bibinfo{journal}{Science} \textbf{\bibinfo{volume}{320}},
  \bibinfo{pages}{899} (\bibinfo{year}{2008}).

\bibitem[{\citenamefont{Willett et~al.}(2009)\citenamefont{Willett, Pfeiffer,
  and West}}]{Willet_PNAS09}
\bibinfo{author}{\bibfnamefont{R.~L.} \bibnamefont{Willett}},
  \bibinfo{author}{\bibfnamefont{L.~N.} \bibnamefont{Pfeiffer}},
  \bibnamefont{and} \bibinfo{author}{\bibfnamefont{K.~W.} \bibnamefont{West}},
  \bibinfo{journal}{Proceedings of the National Academy of Sciences}
  \textbf{\bibinfo{volume}{106}}, \bibinfo{pages}{8853} (\bibinfo{year}{2009}).

\bibitem[{\citenamefont{Bishara et~al.}(2009)\citenamefont{Bishara, Bonderson,
  Nayak, Shtengel, and Slingerland}}]{Nayak_PRB'09}
\bibinfo{author}{\bibfnamefont{W.}~\bibnamefont{Bishara}},
  \bibinfo{author}{\bibfnamefont{P.}~\bibnamefont{Bonderson}},
  \bibinfo{author}{\bibfnamefont{C.}~\bibnamefont{Nayak}},
  \bibinfo{author}{\bibfnamefont{K.}~\bibnamefont{Shtengel}}, \bibnamefont{and}
  \bibinfo{author}{\bibfnamefont{J.~K.} \bibnamefont{Slingerland}},
  \bibinfo{journal}{Phys. Rev. B} \textbf{\bibinfo{volume}{80}},
  \bibinfo{pages}{155303} (\bibinfo{year}{2009}).

\bibitem[{\citenamefont{Kopnin and Salomaa}(1991)}]{Kopnin_PRB'91}
\bibinfo{author}{\bibfnamefont{N.~B.} \bibnamefont{Kopnin}} \bibnamefont{and}
  \bibinfo{author}{\bibfnamefont{M.~M.} \bibnamefont{Salomaa}},
  \bibinfo{journal}{Phys. Rev. B} \textbf{\bibinfo{volume}{44}},
  \bibinfo{pages}{9667} (\bibinfo{year}{1991}).

\bibitem[{\citenamefont{Tsutsumi et~al.}(2008)\citenamefont{Tsutsumi, Kawakami,
  Mizushima, Ichioka, and Machida}}]{Tsutsumi_PRL08}
\bibinfo{author}{\bibfnamefont{Y.}~\bibnamefont{Tsutsumi}},
  \bibinfo{author}{\bibfnamefont{T.}~\bibnamefont{Kawakami}},
  \bibinfo{author}{\bibfnamefont{T.}~\bibnamefont{Mizushima}},
  \bibinfo{author}{\bibfnamefont{M.}~\bibnamefont{Ichioka}}, \bibnamefont{and}
  \bibinfo{author}{\bibfnamefont{K.}~\bibnamefont{Machida}},
  \bibinfo{journal}{Phys. Rev. Lett.} \textbf{\bibinfo{volume}{101}},
  \bibinfo{pages}{135302} (\bibinfo{year}{2008}).

\bibitem[{\citenamefont{Mackenzie and Maeno}(2003)}]{Mackenzie_RevModPhys'03}
\bibinfo{author}{\bibfnamefont{A.~P.} \bibnamefont{Mackenzie}}
  \bibnamefont{and} \bibinfo{author}{\bibfnamefont{Y.}~\bibnamefont{Maeno}},
  \bibinfo{journal}{Rev. Mod. Phys.} \textbf{\bibinfo{volume}{75}},
  \bibinfo{pages}{657} (\bibinfo{year}{2003}).

\bibitem[{\citenamefont{Xia et~al.}(2006)\citenamefont{Xia, Maeno, Beyersdorf,
  Fejer, and Kapitulnik}}]{Kapitulnik_PRL'06}
\bibinfo{author}{\bibfnamefont{J.}~\bibnamefont{Xia}},
  \bibinfo{author}{\bibfnamefont{Y.}~\bibnamefont{Maeno}},
  \bibinfo{author}{\bibfnamefont{P.~T.} \bibnamefont{Beyersdorf}},
  \bibinfo{author}{\bibfnamefont{M.~M.} \bibnamefont{Fejer}}, \bibnamefont{and}
  \bibinfo{author}{\bibfnamefont{A.}~\bibnamefont{Kapitulnik}},
  \bibinfo{journal}{Phys. Rev. Lett.} \textbf{\bibinfo{volume}{97}},
  \bibinfo{pages}{167002} (\bibinfo{year}{2006}).

\bibitem[{\citenamefont{Lutchyn et~al.}(2008)\citenamefont{Lutchyn, Nagornykh,
  and Yakovenko}}]{lutchyn_prb'08}
\bibinfo{author}{\bibfnamefont{R.~M.} \bibnamefont{Lutchyn}},
  \bibinfo{author}{\bibfnamefont{P.}~\bibnamefont{Nagornykh}},
  \bibnamefont{and} \bibinfo{author}{\bibfnamefont{V.~M.}
  \bibnamefont{Yakovenko}}, \bibinfo{journal}{Phys.\ Rev.\ B}
  \textbf{\bibinfo{volume}{77}}, \bibinfo{eid}{144516} (\bibinfo{year}{2008}).

\bibitem[{\citenamefont{Lutchyn et~al.}(2009)\citenamefont{Lutchyn, Nagornykh,
  and Yakovenko}}]{Lutchyn_prb'09}
\bibinfo{author}{\bibfnamefont{R.~M.} \bibnamefont{Lutchyn}},
  \bibinfo{author}{\bibfnamefont{P.}~\bibnamefont{Nagornykh}},
  \bibnamefont{and} \bibinfo{author}{\bibfnamefont{V.~M.}
  \bibnamefont{Yakovenko}}, \bibinfo{journal}{Phys. Rev. B}
  \textbf{\bibinfo{volume}{80}}, \bibinfo{pages}{104508}
  (\bibinfo{year}{2009}).

\bibitem[{\citenamefont{Kallin and Berlinsky}(2009)}]{Kallin}
\bibinfo{author}{\bibfnamefont{C.}~\bibnamefont{Kallin}} \bibnamefont{and}
  \bibinfo{author}{\bibfnamefont{A.~J.} \bibnamefont{Berlinsky}},
  \bibinfo{journal}{Journal of Physics: Condensed Matter}
  \textbf{\bibinfo{volume}{21}}, \bibinfo{pages}{164210}
  (\bibinfo{year}{2009}).

\bibitem[{\citenamefont{Das~Sarma et~al.}(2006)\citenamefont{Das~Sarma, Nayak,
  and Tewari}}]{SDS_PRB06}
\bibinfo{author}{\bibfnamefont{S.}~\bibnamefont{Das~Sarma}},
  \bibinfo{author}{\bibfnamefont{C.}~\bibnamefont{Nayak}}, \bibnamefont{and}
  \bibinfo{author}{\bibfnamefont{S.}~\bibnamefont{Tewari}},
  \bibinfo{journal}{Phys. Rev. B} \textbf{\bibinfo{volume}{73}},
  \bibinfo{pages}{220502} (\bibinfo{year}{2006}).

\bibitem[{\citenamefont{Gurarie and
  Radzihovsky}(2007{\natexlab{a}})}]{Gurarie20072}
\bibinfo{author}{\bibfnamefont{V.}~\bibnamefont{Gurarie}} \bibnamefont{and}
  \bibinfo{author}{\bibfnamefont{L.}~\bibnamefont{Radzihovsky}},
  \bibinfo{journal}{Ann. Phys.(Leipzig)} \textbf{\bibinfo{volume}{322}},
  \bibinfo{pages}{2} (\bibinfo{year}{2007}{\natexlab{a}}).

\bibitem[{\citenamefont{Mizushima et~al.}(2008)\citenamefont{Mizushima,
  Ichioka, and Machida}}]{Mizushima_PRL08}
\bibinfo{author}{\bibfnamefont{T.}~\bibnamefont{Mizushima}},
  \bibinfo{author}{\bibfnamefont{M.}~\bibnamefont{Ichioka}}, \bibnamefont{and}
  \bibinfo{author}{\bibfnamefont{K.}~\bibnamefont{Machida}},
  \bibinfo{journal}{Phys. Rev. Lett.} \textbf{\bibinfo{volume}{101}},
  \bibinfo{pages}{150409} (\bibinfo{year}{2008}).

\bibitem[{\citenamefont{Zhang et~al.}(2008)\citenamefont{Zhang, Tewari,
  Lutchyn, and Das~Sarma}}]{zhang_prl'08}
\bibinfo{author}{\bibfnamefont{C.}~\bibnamefont{Zhang}},
  \bibinfo{author}{\bibfnamefont{S.}~\bibnamefont{Tewari}},
  \bibinfo{author}{\bibfnamefont{R.~M.} \bibnamefont{Lutchyn}},
  \bibnamefont{and}
  \bibinfo{author}{\bibfnamefont{S.}~\bibnamefont{Das~Sarma}},
  \bibinfo{journal}{Phys. Rev. Lett.} \textbf{\bibinfo{volume}{101}},
  \bibinfo{pages}{160401} (\bibinfo{year}{2008}).

\bibitem[{\citenamefont{Sato and Fujimoto}(2009)}]{Sato_PRB09}
\bibinfo{author}{\bibfnamefont{M.}~\bibnamefont{Sato}} \bibnamefont{and}
  \bibinfo{author}{\bibfnamefont{S.}~\bibnamefont{Fujimoto}},
  \bibinfo{journal}{Phys. Rev. B} \textbf{\bibinfo{volume}{79}},
  \bibinfo{pages}{094504} (\bibinfo{year}{2009}).

\bibitem[{\citenamefont{Nishida}(2009)}]{Nishida_AnnPhys09}
\bibinfo{author}{\bibfnamefont{Y.}~\bibnamefont{Nishida}},
  \bibinfo{journal}{Ann. Phys.(N.Y.)} \textbf{\bibinfo{volume}{324}},
  \bibinfo{pages}{897} (\bibinfo{year}{2009}).

\bibitem[{\citenamefont{Cooper and Shlyapnikov}(2009)}]{Cooper_prl'09}
\bibinfo{author}{\bibfnamefont{N.~R.} \bibnamefont{Cooper}} \bibnamefont{and}
  \bibinfo{author}{\bibfnamefont{G.~V.} \bibnamefont{Shlyapnikov}},
  \bibinfo{journal}{Phys. Rev. Lett.} \textbf{\bibinfo{volume}{103}},
  \bibinfo{pages}{155302} (\bibinfo{year}{2009}).

\bibitem[{\citenamefont{Fu and Kane}(2008)}]{Fu_PRL08}
\bibinfo{author}{\bibfnamefont{L.}~\bibnamefont{Fu}} \bibnamefont{and}
  \bibinfo{author}{\bibfnamefont{C.~L.} \bibnamefont{Kane}},
  \bibinfo{journal}{Phys. Rev. Lett.} \textbf{\bibinfo{volume}{100}},
  \bibinfo{pages}{096407} (\bibinfo{year}{2008}).

\bibitem[{\citenamefont{Linder et~al.}(2010)\citenamefont{Linder, Tanaka,
  Yokoyama, Sudb\o{}, and Nagaosa}}]{Linder_PRL10}
\bibinfo{author}{\bibfnamefont{J.}~\bibnamefont{Linder}},
  \bibinfo{author}{\bibfnamefont{Y.}~\bibnamefont{Tanaka}},
  \bibinfo{author}{\bibfnamefont{T.}~\bibnamefont{Yokoyama}},
  \bibinfo{author}{\bibfnamefont{A.}~\bibnamefont{Sudb{\o}}}, \bibnamefont{and}
  \bibinfo{author}{\bibfnamefont{N.}~\bibnamefont{Nagaosa}},
  \bibinfo{journal}{Phys. Rev. Lett.} \textbf{\bibinfo{volume}{104}},
  \bibinfo{pages}{067001} (\bibinfo{year}{2010}).

\bibitem[{\citenamefont{Sau et~al.}(2010)\citenamefont{Sau, Lutchyn, Tewari,
  and Das~Sarma}}]{Sau_PRL10}
\bibinfo{author}{\bibfnamefont{J.~D.} \bibnamefont{Sau}},
  \bibinfo{author}{\bibfnamefont{R.~M.} \bibnamefont{Lutchyn}},
  \bibinfo{author}{\bibfnamefont{S.}~\bibnamefont{Tewari}}, \bibnamefont{and}
  \bibinfo{author}{\bibfnamefont{S.}~\bibnamefont{Das~Sarma}},
  \bibinfo{journal}{Phys. Rev. Lett.} \textbf{\bibinfo{volume}{104}},
  \bibinfo{pages}{040502} (\bibinfo{year}{2010}).

\bibitem[{\citenamefont{Alicea}(2010)}]{Alicea_PRB10}
\bibinfo{author}{\bibfnamefont{J.}~\bibnamefont{Alicea}},
  \bibinfo{journal}{Phys. Rev. B} \textbf{\bibinfo{volume}{81}},
  \bibinfo{pages}{125318} (\bibinfo{year}{2010}).

\bibitem[{\citenamefont{Lee}(2009)}]{Lee_preprint09}
\bibinfo{author}{\bibfnamefont{P.~A.} \bibnamefont{Lee}},
  \bibinfo{journal}{arXiv:0907.2681}  (\bibinfo{year}{2009}).

\bibitem[{\citenamefont{Qi et~al.}(2010)\citenamefont{Qi, Hughes, and
  Zhang}}]{Qi'10}
\bibinfo{author}{\bibfnamefont{X.-L.} \bibnamefont{Qi}},
  \bibinfo{author}{\bibfnamefont{T.~L.} \bibnamefont{Hughes}},
  \bibnamefont{and} \bibinfo{author}{\bibfnamefont{S.-C.} \bibnamefont{Zhang}},
  \bibinfo{journal}{arXiv:1003.5448}  (\bibinfo{year}{2010}).

\bibitem[{\citenamefont{Kitaev}(2001)}]{Kitaev_Majorana}
\bibinfo{author}{\bibfnamefont{A.}~\bibnamefont{Kitaev}},
  \bibinfo{journal}{Physics-Uspekhi} \textbf{\bibinfo{volume}{44}},
  \bibinfo{pages}{131} (\bibinfo{year}{2001}).

\bibitem[{\citenamefont{Fu and Kane}(2009)}]{Fu_PRB2009}
\bibinfo{author}{\bibfnamefont{L.}~\bibnamefont{Fu}} \bibnamefont{and}
  \bibinfo{author}{\bibfnamefont{C.~L.} \bibnamefont{Kane}},
  \bibinfo{journal}{Phys. Rev. B} \textbf{\bibinfo{volume}{79}},
  \bibinfo{pages}{161408} (\bibinfo{year}{2009}).

\bibitem[{\citenamefont{{Wimmer} et~al.}(2010)\citenamefont{{Wimmer},
  {Akhmerov}, {Medvedyeva}, {Tworzyd{\l}o}, and {Beenakker}}}]{Wimmer_arxiv10}
\bibinfo{author}{\bibfnamefont{M.}~\bibnamefont{{Wimmer}}},
  \bibinfo{author}{\bibfnamefont{A.~R.} \bibnamefont{{Akhmerov}}},
  \bibinfo{author}{\bibfnamefont{M.~V.} \bibnamefont{{Medvedyeva}}},
  \bibinfo{author}{\bibfnamefont{J.}~\bibnamefont{{Tworzyd{\l}o}}},
  \bibnamefont{and} \bibinfo{author}{\bibfnamefont{C.~W.~J.}
  \bibnamefont{{Beenakker}}}, \bibinfo{journal}{arXiv:1002.3570}
  (\bibinfo{year}{2010}).

\bibitem[{\citenamefont{Lutchyn et~al.}(2010)\citenamefont{Lutchyn, Sau, and
  Sarma}}]{Lutchyn_preprint10}
\bibinfo{author}{\bibfnamefont{R.~M.} \bibnamefont{Lutchyn}},
  \bibinfo{author}{\bibfnamefont{J.~D.} \bibnamefont{Sau}}, \bibnamefont{and}
  \bibinfo{author}{\bibfnamefont{S.~D.} \bibnamefont{Sarma}},
  \bibinfo{journal}{arXiv:1002.4033}  (\bibinfo{year}{2010}).

\bibitem[{\citenamefont{Oreg et~al.}(2010)\citenamefont{Oreg, Refael, and von
  Oppen}}]{Oreg_2010}
\bibinfo{author}{\bibfnamefont{Y.}~\bibnamefont{Oreg}},
  \bibinfo{author}{\bibfnamefont{G.}~\bibnamefont{Refael}}, \bibnamefont{and}
  \bibinfo{author}{\bibfnamefont{F.}~\bibnamefont{von Oppen}},
  \bibinfo{journal}{arXiv:1003.1145}  (\bibinfo{year}{2010}).

\bibitem[{\citenamefont{Roy}(2008)}]{Roy_preprint08}
\bibinfo{author}{\bibfnamefont{R.}~\bibnamefont{Roy}},
  \bibinfo{journal}{arXiv:0803.2868}  (\bibinfo{year}{2008}).

\bibitem[{\citenamefont{Chung and Zhang}(2009)}]{Chung_PRL09}
\bibinfo{author}{\bibfnamefont{S.~B.} \bibnamefont{Chung}} \bibnamefont{and}
  \bibinfo{author}{\bibfnamefont{S.-C.} \bibnamefont{Zhang}},
  \bibinfo{journal}{Phys. Rev. Lett.} \textbf{\bibinfo{volume}{103}},
  \bibinfo{pages}{235301} (\bibinfo{year}{2009}).

    \bibitem{Volovik_JETP09} G.E. Volovik, JETP Lett. 90, 398 (2009)

\bibitem[{\citenamefont{Qi et~al.}(2009)\citenamefont{Qi, Hughes, Raghu, and
  Zhang}}]{Qi_PRL09}
\bibinfo{author}{\bibfnamefont{X.-L.} \bibnamefont{Qi}},
  \bibinfo{author}{\bibfnamefont{T.~L.} \bibnamefont{Hughes}},
  \bibinfo{author}{\bibfnamefont{S.}~\bibnamefont{Raghu}}, \bibnamefont{and}
  \bibinfo{author}{\bibfnamefont{S.-C.} \bibnamefont{Zhang}},
  \bibinfo{journal}{Phys. Rev. Lett.} \textbf{\bibinfo{volume}{102}},
  \bibinfo{pages}{187001} (\bibinfo{year}{2009}).

\bibitem[{\citenamefont{Hooft and Bruckmann}()}]{tHooft}
\bibinfo{author}{\bibfnamefont{G.}~\bibnamefont{Hooft}} \bibnamefont{and}
  \bibinfo{author}{\bibfnamefont{F.}~\bibnamefont{Bruckmann}},
  \bibinfo{howpublished}{Lectures given at the 5th WE Heraeus Summer School,
  Saalburg/Germany, September 1999, on ``Monopoles, Instantons and
  Confinement,'' hep-th/0010225 (2000)}.

\bibitem[{\citenamefont{Jackiw and Rossi}(1981)}]{Jackiw_NPB81}
\bibinfo{author}{\bibfnamefont{R.}~\bibnamefont{Jackiw}} \bibnamefont{and}
  \bibinfo{author}{\bibfnamefont{P.}~\bibnamefont{Rossi}},
  \bibinfo{journal}{Nucl. Phys. B} \textbf{\bibinfo{volume}{190}},
  \bibinfo{pages}{681 } (\bibinfo{year}{1981}).

\bibitem[{\citenamefont{Weinberg}(1981)}]{Weinberg_PRD81}
\bibinfo{author}{\bibfnamefont{E.~J.} \bibnamefont{Weinberg}},
  \bibinfo{journal}{Phys. Rev. D} \textbf{\bibinfo{volume}{24}},
  \bibinfo{pages}{2669} (\bibinfo{year}{1981}).

\bibitem[{\citenamefont{Atiyah and Singer}(1968)}]{Atiyah_AnnMath68}
\bibinfo{author}{\bibfnamefont{M.~F.} \bibnamefont{Atiyah}} \bibnamefont{and}
  \bibinfo{author}{\bibfnamefont{I.~M.} \bibnamefont{Singer}},
  \bibinfo{journal}{The Annals of Mathematics} \textbf{\bibinfo{volume}{87}},
  \bibinfo{pages}{484} (\bibinfo{year}{1968}).

\bibitem[{Jac()}]{Jackiw_private}
\bibinfo{note}{Strictly speaking, Atiyah-Singer theorem relates the number
  difference of zero modes of opposite chiralities with the topological index
  defined in terms of the gauge field strength tensor. In our case the
  topological index is given in terms of the winding number of the scalar
  field. We thank R. Jackiw for pointing this out.}

\bibitem[{\citenamefont{Chamon et~al.}(2010)\citenamefont{Chamon, Jackiw,
  Nishida, Pi, and Santos}}]{Chamon_PRB10}
\bibinfo{author}{\bibfnamefont{C.}~\bibnamefont{Chamon}},
  \bibinfo{author}{\bibfnamefont{R.}~\bibnamefont{Jackiw}},
  \bibinfo{author}{\bibfnamefont{Y.}~\bibnamefont{Nishida}},
  \bibinfo{author}{\bibfnamefont{S.-Y.} \bibnamefont{Pi}}, \bibnamefont{and}
  \bibinfo{author}{\bibfnamefont{L.}~\bibnamefont{Santos}},
  \bibinfo{journal}{Phys. Rev. B} \textbf{\bibinfo{volume}{81}},
  \bibinfo{pages}{224515} (\bibinfo{year}{2010}).

\bibitem[{\citenamefont{Bonderson}(2009)}]{Bonderson_PRL09}
\bibinfo{author}{\bibfnamefont{P.}~\bibnamefont{Bonderson}},
  \bibinfo{journal}{Phys. Rev. Lett.} \textbf{\bibinfo{volume}{103}},
  \bibinfo{pages}{110403} (\bibinfo{year}{2009}).

\bibitem[{\citenamefont{Tserkovnyak and Simon}(2003)}]{Tserkovnyak_PRL03}
\bibinfo{author}{\bibfnamefont{Y.}~\bibnamefont{Tserkovnyak}} \bibnamefont{and}
  \bibinfo{author}{\bibfnamefont{S.~H.} \bibnamefont{Simon}},
  \bibinfo{journal}{Phys. Rev. Lett.} \textbf{\bibinfo{volume}{90}},
  \bibinfo{pages}{016802} (\bibinfo{year}{2003}).

\bibitem[{\citenamefont{Baraban et~al.}(2009)\citenamefont{Baraban, Zikos,
  Bonesteel, and Simon}}]{Baraban_PRL09}
\bibinfo{author}{\bibfnamefont{M.}~\bibnamefont{Baraban}},
  \bibinfo{author}{\bibfnamefont{G.}~\bibnamefont{Zikos}},
  \bibinfo{author}{\bibfnamefont{N.}~\bibnamefont{Bonesteel}},
  \bibnamefont{and} \bibinfo{author}{\bibfnamefont{S.~H.} \bibnamefont{Simon}},
  \bibinfo{journal}{Phys. Rev. Lett.} \textbf{\bibinfo{volume}{103}},
  \bibinfo{pages}{076801} (\bibinfo{year}{2009}).

\bibitem[{\citenamefont{Kraus et~al.}(2009)\citenamefont{Kraus, Auerbach,
  Fertig, and Simon}}]{Kraus_PRB09}
\bibinfo{author}{\bibfnamefont{Y.~E.} \bibnamefont{Kraus}},
  \bibinfo{author}{\bibfnamefont{A.}~\bibnamefont{Auerbach}},
  \bibinfo{author}{\bibfnamefont{H.~A.} \bibnamefont{Fertig}},
  \bibnamefont{and} \bibinfo{author}{\bibfnamefont{S.~H.} \bibnamefont{Simon}},
  \bibinfo{journal}{Phys. Rev. B} \textbf{\bibinfo{volume}{79}},
  \bibinfo{pages}{134515} (\bibinfo{year}{2009}).

\bibitem[{\citenamefont{Lahtinen et~al.}(2008)\citenamefont{Lahtinen, Kells,
  Carollo, Stitt, Vala, and Pachos}}]{Lahtinen_AnnPhys08}
\bibinfo{author}{\bibfnamefont{V.}~\bibnamefont{Lahtinen}},
  \bibinfo{author}{\bibfnamefont{G.}~\bibnamefont{Kells}},
  \bibinfo{author}{\bibfnamefont{A.}~\bibnamefont{Carollo}},
  \bibinfo{author}{\bibfnamefont{T.}~\bibnamefont{Stitt}},
  \bibinfo{author}{\bibfnamefont{J.}~\bibnamefont{Vala}}, \bibnamefont{and}
  \bibinfo{author}{\bibfnamefont{J.~K.} \bibnamefont{Pachos}},
  \bibinfo{journal}{Ann. Phys.(N.Y.)} \textbf{\bibinfo{volume}{323}},
  \bibinfo{pages}{2286 } (\bibinfo{year}{2008}).

\bibitem[{\citenamefont{Cheng et~al.}(2009)\citenamefont{Cheng, Lutchyn,
  Galitski, and Das~Sarma}}]{Cheng_PRL09}
\bibinfo{author}{\bibfnamefont{M.}~\bibnamefont{Cheng}},
  \bibinfo{author}{\bibfnamefont{R.~M.} \bibnamefont{Lutchyn}},
  \bibinfo{author}{\bibfnamefont{V.}~\bibnamefont{Galitski}}, \bibnamefont{and}
  \bibinfo{author}{\bibfnamefont{S.}~\bibnamefont{Das~Sarma}},
  \bibinfo{journal}{Phys. Rev. Lett.} \textbf{\bibinfo{volume}{103}},
  \bibinfo{pages}{107001} (\bibinfo{year}{2009}).

\bibitem[{\citenamefont{Stern et~al.}(2004)\citenamefont{Stern, von Oppen, and
  Mariani}}]{Stern_prb'04}
\bibinfo{author}{\bibfnamefont{A.}~\bibnamefont{Stern}},
  \bibinfo{author}{\bibfnamefont{F.}~\bibnamefont{von Oppen}},
  \bibnamefont{and} \bibinfo{author}{\bibfnamefont{E.}~\bibnamefont{Mariani}},
  \bibinfo{journal}{Phys. Rev. B} \textbf{\bibinfo{volume}{70}},
  \bibinfo{pages}{205338} (\bibinfo{year}{2004}).

\bibitem[{\citenamefont{Cheng et~al.}(2010)\citenamefont{Cheng, Sun, Galitski,
  and Das~Sarma}}]{Cheng_PRB2010}
\bibinfo{author}{\bibfnamefont{M.}~\bibnamefont{Cheng}},
  \bibinfo{author}{\bibfnamefont{K.}~\bibnamefont{Sun}},
  \bibinfo{author}{\bibfnamefont{V.}~\bibnamefont{Galitski}}, \bibnamefont{and}
  \bibinfo{author}{\bibfnamefont{S.}~\bibnamefont{Das~Sarma}},
  \bibinfo{journal}{Phys. Rev. B} \textbf{\bibinfo{volume}{81}},
  \bibinfo{pages}{024504} (\bibinfo{year}{2010}).

\bibitem[{\citenamefont{Gurarie and
  Radzihovsky}(2007{\natexlab{b}})}]{Gurarie_prb'07}
\bibinfo{author}{\bibfnamefont{V.}~\bibnamefont{Gurarie}} \bibnamefont{and}
  \bibinfo{author}{\bibfnamefont{L.}~\bibnamefont{Radzihovsky}},
  \bibinfo{journal}{Phys.\ Rev.\ B} \textbf{\bibinfo{volume}{75}},
  \bibinfo{pages}{212509} (\bibinfo{year}{2007}{\natexlab{b}}).

\bibitem[{\citenamefont{Schnyder et~al.}(2008)\citenamefont{Schnyder, Ryu,
  Furusaki, and Ludwig}}]{Schnyder_PRB08}
\bibinfo{author}{\bibfnamefont{A.~P.} \bibnamefont{Schnyder}},
  \bibinfo{author}{\bibfnamefont{S.}~\bibnamefont{Ryu}},
  \bibinfo{author}{\bibfnamefont{A.}~\bibnamefont{Furusaki}}, \bibnamefont{and}
  \bibinfo{author}{\bibfnamefont{A.~W.~W.} \bibnamefont{Ludwig}},
  \bibinfo{journal}{Phys. Rev. B} \textbf{\bibinfo{volume}{78}},
  \bibinfo{pages}{195125} (\bibinfo{year}{2008}).

\bibitem[{\citenamefont{Kitaev}(2009)}]{Kitaev2009}
\bibinfo{author}{\bibfnamefont{A.}~\bibnamefont{Kitaev}}, \emph{\bibinfo{title}{Proceedings of the L.D. Landau Memorial Conference
Advances in Theoretical Physics}},
  \bibinfo{publisher}{Chernogolovka, Moscow region, Russia}, 22-26 June 2008 (unpublished)

\bibitem[{\citenamefont{Caroli et~al.}(1964)\citenamefont{Caroli, de~Gennes,
  and Matricon}}]{Caroli_PL'64}
\bibinfo{author}{\bibfnamefont{C.}~\bibnamefont{Caroli}},
  \bibinfo{author}{\bibfnamefont{P.}~\bibnamefont{de~Gennes}},
  \bibnamefont{and} \bibinfo{author}{\bibfnamefont{J.}~\bibnamefont{Matricon}},
  \bibinfo{journal}{Phys. Lett.} \textbf{\bibinfo{volume}{9}},
  \bibinfo{pages}{307} (\bibinfo{year}{1964}).

\bibitem[{\citenamefont{Tewari et~al.}(2007{\natexlab{b}})\citenamefont{Tewari,
  Das~Sarma, and Lee}}]{tewari_prl'07}
\bibinfo{author}{\bibfnamefont{S.}~\bibnamefont{Tewari}},
  \bibinfo{author}{\bibfnamefont{S.} \bibnamefont{Das~Sarma}}, \bibnamefont{and}
  \bibinfo{author}{\bibfnamefont{D.-H.} \bibnamefont{Lee}},
  \bibinfo{journal}{Phys.\ Rev.\ Lett.} \textbf{\bibinfo{volume}{99}},
  \bibinfo{pages}{037001} (\bibinfo{year}{2007}{\natexlab{b}}).

\bibitem[{\citenamefont{Mizushima and
  Machida}(2010{\natexlab{a}})}]{Mizushima_PRA10}
\bibinfo{author}{\bibfnamefont{T.}~\bibnamefont{Mizushima}} \bibnamefont{and}
  \bibinfo{author}{\bibfnamefont{K.}~\bibnamefont{Machida}},
  \bibinfo{journal}{Phys. Rev. A} \textbf{\bibinfo{volume}{81}},
  \bibinfo{pages}{053605} (\bibinfo{year}{2010}{\natexlab{a}}).

\bibitem[{\citenamefont{Bardeen et~al.}(1969)\citenamefont{Bardeen, K\"ummel,
  Jacobs, and Tewordt}}]{bardeen_prb'69}
\bibinfo{author}{\bibfnamefont{J.}~\bibnamefont{Bardeen}},
  \bibinfo{author}{\bibfnamefont{R.}~\bibnamefont{K\"ummel}},
  \bibinfo{author}{\bibfnamefont{A.~E.} \bibnamefont{Jacobs}},
  \bibnamefont{and} \bibinfo{author}{\bibfnamefont{L.}~\bibnamefont{Tewordt}},
  \bibinfo{journal}{Phys. Rev.} \textbf{\bibinfo{volume}{187}},
  \bibinfo{pages}{556} (\bibinfo{year}{1969}).

\bibitem[{\citenamefont{Fu et~al.}(2007)\citenamefont{Fu, Kane, and
  Mele}}]{Fu_PRL07}
\bibinfo{author}{\bibfnamefont{L.}~\bibnamefont{Fu}},
  \bibinfo{author}{\bibfnamefont{C.~L.} \bibnamefont{Kane}}, \bibnamefont{and}
  \bibinfo{author}{\bibfnamefont{E.~J.} \bibnamefont{Mele}},
  \bibinfo{journal}{Phys. Rev. Lett.} \textbf{\bibinfo{volume}{98}},
  \bibinfo{pages}{106803} (\bibinfo{year}{2007}).

\bibitem[{\citenamefont{Roy}(2009)}]{Roy_PRB09}
\bibinfo{author}{\bibfnamefont{R.}~\bibnamefont{Roy}}, \bibinfo{journal}{Phys.
  Rev. B} \textbf{\bibinfo{volume}{79}}, \bibinfo{pages}{195322}
  (\bibinfo{year}{2009}).

\bibitem[{\citenamefont{Moore and Balents}(2007)}]{Moore_PRB07}
\bibinfo{author}{\bibfnamefont{J.~E.} \bibnamefont{Moore}} \bibnamefont{and}
  \bibinfo{author}{\bibfnamefont{L.}~\bibnamefont{Balents}},
  \bibinfo{journal}{Phys. Rev. B} \textbf{\bibinfo{volume}{75}},
  \bibinfo{pages}{121306} (\bibinfo{year}{2007}).

\bibitem[{\citenamefont{{Stanescu} et~al.}(2010)\citenamefont{{Stanescu},
  {Sau}, {Lutchyn}, and {Das Sarma}}}]{Stanescu'10}
\bibinfo{author}{\bibfnamefont{T.~D.} \bibnamefont{{Stanescu}}},
  \bibinfo{author}{\bibfnamefont{J.~D.} \bibnamefont{{Sau}}},
  \bibinfo{author}{\bibfnamefont{R.~M.} \bibnamefont{{Lutchyn}}},
  \bibnamefont{and} \bibinfo{author}{\bibfnamefont{S.}~\bibnamefont{{Das
  Sarma}}}, \bibinfo{journal}{Phys. Rev. B} \textbf{\bibinfo{volume}{81}},
  \bibinfo{pages}{241310} (\bibinfo{year}{2010}).

\bibitem[{\citenamefont{{Ludwig} et~al.}(2010)\citenamefont{{Ludwig},
  {Poilblanc}, {Trebst}, and {Troyer}}}]{Ludwig_arxiv'10}
\bibinfo{author}{\bibfnamefont{A.~W.~W.} \bibnamefont{{Ludwig}}},
  \bibinfo{author}{\bibfnamefont{D.}~\bibnamefont{{Poilblanc}}},
  \bibinfo{author}{\bibfnamefont{S.}~\bibnamefont{{Trebst}}}, \bibnamefont{and}
  \bibinfo{author}{\bibfnamefont{M.}~\bibnamefont{{Troyer}}},
  \bibinfo{journal}{arXiv:1003.3453}  (\bibinfo{year}{2010}).

\bibitem[{\citenamefont{Landau and Lifshitz}(1984)}]{Landau_book3}
\bibinfo{author}{\bibfnamefont{L.}~\bibnamefont{Landau}} \bibnamefont{and}
  \bibinfo{author}{\bibfnamefont{E.}~\bibnamefont{Lifshitz}},
  \emph{\bibinfo{title}{Quantum Mechanics (Course of Theoretical Physics)}},
  vol.~\bibinfo{volume}{3} (\bibinfo{publisher}{Pergamon Press},
  \bibinfo{year}{1984}), \bibinfo{edition}{2nd} ed.

\bibitem[{\citenamefont{Galitsky and Sokoloff}()}]{VGMHD}
\bibinfo{author}{\bibfnamefont{V.~M.} \bibnamefont{Galitsky}} \bibnamefont{and}
  \bibinfo{author}{\bibfnamefont{D.~D.} \bibnamefont{Sokoloff}},
  \bibinfo{howpublished}{Geophys. Astrophys. Fluid Dynamics {\bf 91}, 147
  (1999); V. M. Galitski, K. M. Kuzanyan, and D. D. Sokoloff, Astron.Rep. {\bf
  49}, 337 (2005)}.

\bibitem[{\citenamefont{Nilsson et~al.}(2008)\citenamefont{Nilsson, Akhmerov,
  and Beenakker}}]{Nilsson_PRL08}
\bibinfo{author}{\bibfnamefont{J.}~\bibnamefont{Nilsson}},
  \bibinfo{author}{\bibfnamefont{A.~R.} \bibnamefont{Akhmerov}},
  \bibnamefont{and} \bibinfo{author}{\bibfnamefont{C.~W.~J.}
  \bibnamefont{Beenakker}}, \bibinfo{journal}{Phys. Rev. Lett.}
  \textbf{\bibinfo{volume}{101}}, \bibinfo{pages}{120403}
  (\bibinfo{year}{2008}).

\bibitem[{\citenamefont{Fukui and Fujiwara}(2010)}]{Fukui_JPSJ10}
\bibinfo{author}{\bibfnamefont{T.}~\bibnamefont{Fukui}} \bibnamefont{and}
  \bibinfo{author}{\bibfnamefont{T.}~\bibnamefont{Fujiwara}},
  \bibinfo{journal}{J. Phys. Soc. Jpn.} \textbf{\bibinfo{volume}{79}},
  \bibinfo{pages}{033701} (\bibinfo{year}{2010}).

\bibitem[{\citenamefont{Grosfeld and Stern}(2006)}]{Grosfeld_prb'06}
\bibinfo{author}{\bibfnamefont{E.}~\bibnamefont{Grosfeld}} \bibnamefont{and}
  \bibinfo{author}{\bibfnamefont{A.}~\bibnamefont{Stern}},
  \bibinfo{journal}{Phys. Rev. B} \textbf{\bibinfo{volume}{73}},
  \bibinfo{pages}{201303} (\bibinfo{year}{2006}).

\bibitem[{\citenamefont{Feiguin et~al.}(2007)\citenamefont{Feiguin, Trebst,
  Ludwig, Troyer, Kitaev, Wang, and Freedman}}]{Feiguin_prl'07}
\bibinfo{author}{\bibfnamefont{A.}~\bibnamefont{Feiguin}},
  \bibinfo{author}{\bibfnamefont{S.}~\bibnamefont{Trebst}},
  \bibinfo{author}{\bibfnamefont{A.~W.~W.} \bibnamefont{Ludwig}},
  \bibinfo{author}{\bibfnamefont{M.}~\bibnamefont{Troyer}},
  \bibinfo{author}{\bibfnamefont{A.}~\bibnamefont{Kitaev}},
  \bibinfo{author}{\bibfnamefont{Z.}~\bibnamefont{Wang}}, \bibnamefont{and}
  \bibinfo{author}{\bibfnamefont{M.~H.} \bibnamefont{Freedman}},
  \bibinfo{journal}{Phys. Rev. Lett.} \textbf{\bibinfo{volume}{98}},
  \bibinfo{pages}{160409} (\bibinfo{year}{2007}).

\bibitem[{\citenamefont{Gils et~al.}(2009)\citenamefont{Gils, Ardonne, Trebst,
  Ludwig, Troyer, and Wang}}]{Gils_prl'09}
\bibinfo{author}{\bibfnamefont{C.}~\bibnamefont{Gils}},
  \bibinfo{author}{\bibfnamefont{E.}~\bibnamefont{Ardonne}},
  \bibinfo{author}{\bibfnamefont{S.}~\bibnamefont{Trebst}},
  \bibinfo{author}{\bibfnamefont{A.~W.~W.} \bibnamefont{Ludwig}},
  \bibinfo{author}{\bibfnamefont{M.}~\bibnamefont{Troyer}}, \bibnamefont{and}
  \bibinfo{author}{\bibfnamefont{Z.}~\bibnamefont{Wang}},
  \bibinfo{journal}{Phys. Rev. Lett.} \textbf{\bibinfo{volume}{103}},
  \bibinfo{pages}{070401} (\bibinfo{year}{2009}).

\bibitem[{\citenamefont{{Lahtinen} and {Pachos}}(2010)}]{Lahtinen_PRB'10}
\bibinfo{author}{\bibfnamefont{V.}~\bibnamefont{{Lahtinen}}} \bibnamefont{and}
  \bibinfo{author}{\bibfnamefont{J.~K.} \bibnamefont{{Pachos}}},
  \bibinfo{journal}{Phys. Rev. B} \textbf{\bibinfo{volume}{81}},
  \bibinfo{pages}{245132} (\bibinfo{year}{2010}).

\bibitem[{\citenamefont{Bonderson and Slingerland}(2008)}]{Bonderson_prb'08}
\bibinfo{author}{\bibfnamefont{P.}~\bibnamefont{Bonderson}} \bibnamefont{and}
  \bibinfo{author}{\bibfnamefont{J.~K.} \bibnamefont{Slingerland}},
  \bibinfo{journal}{Phys. Rev. B} \textbf{\bibinfo{volume}{78}},
  \bibinfo{pages}{125323} (\bibinfo{year}{2008}).

\bibitem[{\citenamefont{Levin and Halperin}(2009)}]{Levin_prb'09}
\bibinfo{author}{\bibfnamefont{M.}~\bibnamefont{Levin}} \bibnamefont{and}
  \bibinfo{author}{\bibfnamefont{B.~I.} \bibnamefont{Halperin}},
  \bibinfo{journal}{Phys. Rev. B} \textbf{\bibinfo{volume}{79}},
  \bibinfo{pages}{205301} (\bibinfo{year}{2009}).

\bibitem[{\citenamefont{Hermanns}(2010)}]{Hermanns_prl'10}
\bibinfo{author}{\bibfnamefont{M.}~\bibnamefont{Hermanns}},
  \bibinfo{journal}{Phys. Rev. Lett.} \textbf{\bibinfo{volume}{104}},
  \bibinfo{pages}{056803} (\bibinfo{year}{2010}).

\bibitem[{\citenamefont{Read and Ludwig}(2000)}]{read_prb'00_1}
\bibinfo{author}{\bibfnamefont{N.}~\bibnamefont{Read}} \bibnamefont{and}
  \bibinfo{author}{\bibfnamefont{A.~W.~W.} \bibnamefont{Ludwig}},
  \bibinfo{journal}{Phys. Rev. B} \textbf{\bibinfo{volume}{63}},
  \bibinfo{pages}{024404} (\bibinfo{year}{2000}).

\bibitem[{\citenamefont{Zyuzin and Spivak}()}]{ZSVGAI}
\bibinfo{author}{\bibfnamefont{A.}~\bibnamefont{Zyuzin}} \bibnamefont{and}
  \bibinfo{author}{\bibfnamefont{B.~Z.} \bibnamefont{Spivak}},
  \bibinfo{howpublished}{Pis'ma Zh. Eksp. Teor. Fiz. {\bf 43}, 185 (1986) [JETP
  Lett. {\bf 43}, 234 (1986)]; V. M. Galitski, M. G. Vavilov, and L. I.
  Glazman, Phys. Rev. Lett {\bf 94}, 096602 (2005); V. M. Galitski and A. I.
  Larkin, Phys. Rev. B {\bf 66}, 064526 (2002)}.

\bibitem[{\citenamefont{Mizushima and
  Machida}(2010{\natexlab{b}})}]{Mizushima_arxiv10_5}
\bibinfo{author}{\bibfnamefont{T.}~\bibnamefont{Mizushima}} \bibnamefont{and}
  \bibinfo{author}{\bibfnamefont{K.}~\bibnamefont{Machida}},
  \bibinfo{journal}{arXiv:1005.4738}  (\bibinfo{year}{2010}{\natexlab{b}}).

\end{thebibliography}

\end{document}